\documentclass[preprint,review,12pt]{elsarticle}
\usepackage{graphicx}
\usepackage[percent]{overpic}
\usepackage{amssymb}
\usepackage{amsmath}
\usepackage{epstopdf}
\usepackage{booktabs}
\usepackage{siunitx} 
\newcommand{\Nuc}[2]{\ensuremath{^{#2}\mbox{#1}}}

\journal{Astroparticle Physics}
\begin{document}
\begin{frontmatter}

\author[TRIUMF]{P.-A.~Amaudruz}
\author[LU]{M.~Batygov}
\author[UofA]{B.~Beltran}
\author[Carleton]{K.~Boudjemline}
\author[Queens]{M.G.~Boulay} 
\author[Queens]{B.~Cai} 
\author[UPenn]{T.~Caldwell}
\author[Queens]{M.~Chen}
\author[UofA]{R.~Chouinard}
\author[LU]{B.T.~Cleveland}
\author[SNOLAB]{D.~Contreras}
\author[Queens]{K.~Dering}
\author[LU]{F.~Duncan}
\author[SNOLAB]{R.~Ford}
\author[Queens]{R.~Gagnon}
\author[UNM]{F.~Giuliani}
\author[UNM]{M.~Gold} 
\author[Queens]{V.V.~Golovko\fnref{aecl}}\fntext[aecl]{Current address: AECL, Chalk River, ON, K0J 1J0, Canada.}
\author[UofA]{P.~Gorel}
\author[Carleton]{K.~Graham}
\author[UofA]{D.R.~Grant}
\author[UofA]{R.~Hakobyan}
\author[UofA]{A.L.~Hallin}
\author[Queens]{P.~Harvey}
\author[Queens]{C. Hearns}
\author[LU]{C.J.~Jillings}
\author[Queens]{M.~Ku\'zniak} 
\author[SNOLAB]{I.~Lawson}
\author[SNOLAB]{O.~Li}
\author[Queens]{J.~Lidgard}
\author[SNOLAB]{P.~Liimatainen}
\author[Yale]{W.H.~Lippincott\fnref{fermi}}\fntext[fermi]{Current address: Fermilab, Batavia, IL 60510, USA.}
\author[Queens]{R.~Mathew}
\author[Queens]{A.B.~McDonald}
\author[UofA]{T.~McElroy}
\author[SNOLAB]{K.~McFarlane} 
\author[Yale]{D.~McKinsey}
\author[TRIUMF]{A.~Muir} 
\author[Queens]{C.~Nantais} 
\author[Queens]{K.~Nicolics}
\author[Yale]{J.~Nikkel} 
\author[Queens]{T.~Noble}
\author[Queens]{E.~O'Dwyer} 
\author[UofA]{K.S.~Olsen}
\author[Carleton]{C.~Ouellet}
\author[Queens]{P.~Pasuthip} 
\author[Queens]{T.~Pollmann\corref{cor}\fnref{now}}\cortext[cor]{Corresponding author.}\fntext[now]{Current address: SNOLAB, Lively, ON, P3Y 1N2, Canada.}\ead{tina@snolab.ca}
\author[Queens]{W.~Rau}
\author[TRIUMF]{F.~Retiere}
\author[UNC]{M.~Ronquest} 
\author[Queens]{P.~Skensved} 
\author[Queens]{T.~Sonley} 
\author[UofA]{J.~Tang}
\author[SNOLAB]{E.~V\'azquez-J\'auregui} 
\author[Queens]{L.~Veloce}
\author[Queens]{M.~Ward} 

\address[UofA]{Department of Physics, University of Alberta, Edmonton, AB, T6G 2R3, Canada}
\address[Carleton]{Department of Physics, Carleton University, Ottawa, ON, K1S 5B6, Canada}
\address[LU]{Department of Physics and Astronomy, Laurentian University, Sudbury, ON, P3E 2C6, Canada}
\address[UNM]{Department of Physics, University of New Mexico, Albuquerque, NM 87131, USA}
\address[UNC]{University of North Carolina, Chapel Hill, NC 27517, USA}
\address[UPenn]{Department of Physics, University of Pennsylvania, Philadelphia, PA 19104, USA}
\address[Queens]{Department of Physics, Engineering Physics and Astronomy, Queen's University, Kingston, ON, K7L 3N6, Canada}
\address[SNOLAB]{SNOLAB, Lively, ON, P3Y 1N2, Canada}
\address[TRIUMF]{TRIUMF, Vancouver, BC, V6T 2A3, Canada}
\address[Yale]{Department of Physics, Yale University, New Haven, CT 06511, USA}

\title{Radon backgrounds in the DEAP-1 liquid-argon-based Dark Matter detector}

\begin{abstract}
The DEAP-1 \SI{7}{kg} single phase liquid argon scintillation detector was operated underground at SNOLAB in order to test the techniques and measure the backgrounds inherent to single phase detection, in support of the \mbox{DEAP-3600} Dark Matter detector. Backgrounds in DEAP are controlled through material selection, construction techniques, pulse shape discrimination, and event reconstruction. This report details the analysis of background events observed in three iterations of the DEAP-1 detector, and the measures taken to reduce them.

The $^{222}$Rn decay rate in the liquid argon was measured to be between 16 and \SI{26}{\micro\becquerel\per\kilogram}.
We found that the background spectrum near the region of interest for Dark Matter detection in the DEAP-1 detector can be described considering events from three sources: radon daughters decaying on the surface of the active volume, the expected rate of electromagnetic events misidentified as nuclear recoils due to inefficiencies in the pulse shape discrimination, and leakage of events from outside the fiducial volume due to imperfect position reconstruction. These backgrounds statistically account for all observed events, and they will be strongly reduced in the DEAP-3600 detector due to its higher light yield and simpler geometry.

\end{abstract}

\end{frontmatter}

\section{Organization of this paper}
There are three main parts to this paper. The first part introduces the reader to liquid-argon-based Dark Matter detection (Sect.~\ref{sect:intro}), and the design of the DEAP-1 detector (Sect.~\ref{sect:deap1}). Sources of background (Sect.~\ref{sect:d1background}) and measures taken to reduce or eliminate those sources (Sect.~\ref{sect:bgreduction}), are given special consideration.

The next part outlines the basic data analysis steps and methods (Sect.~\ref{sect:dataanalysis}), followed by a detailed discussion of the populations of nuclear-recoil-events observed in three detector generations (Sect.~\ref{sect:bganalysis}). The ultimate goal of this discussion is to understand the backgrounds in the Dark Matter search window, which means nuclear-recoil-like events at low energies. To arrive at this understanding though, we have to start at the much higher energies of radon-chain alpha decays, and then trace their signatures down to lower energies.

The main tools used to study the decay rates and behaviour of radon chain isotopes are timing coincidence tagging of the radon-polonium and bismuth-polonium pairs (Sect.~\ref{sec:tagging}), and fits to the high energy alpha spectra (Sect.~\ref{subs:rnspectra}).

These tools, supplemented by Monte Carlo simulation, provide insight into the intricate relation between an event's position in the detector and its apparent energy, through effects of macro-geometry at the edges of the detector (Sect. \ref{susbs:geometric}), and micro-geometry at surfaces combined with scintillation of the wavelength-shifter coating (Sect. \ref{sect:surface}).


As a result of our evolving understanding of background sources, (and in response to the need to test components and methods for DEAP-3600), DEAP-1 underwent three major upgrades, so that there is a feedback loop between detector design, background sources and use of analysis methods. Total background rates in the region of interest for Dark Matter search for all three detector generations are given in Sect.~\ref{sect:lowenergyrates}. Limits on the contamination of the detector surfaces with radon chain isotopes for the last detector generation are discussed in Sect.~\ref{sect:surfacebglimits}.

The last part of this paper discusses the details of the background spectrum near the Dark Matter region of interest for the last and cleanest detector generation (Sect.~\ref{sect:lespectrum}), and the discussion (Sect.~\ref{sect:discussion}) touches on implications for DEAP-3600.

\section{Detection of Dark Matter with single phase argon detectors} \label{sect:intro}
Liquid argon is an excellent medium for Dark Matter detection.  It offers large target masses, exceptional purity, efficient scintillation, and pulse shape discrimination (PSD). Several large Dark Matter detectors based on liquid argon have been proposed, including DEAP~\cite{Boulay:2012er}, MiniCLEAN~\cite{miniclean}, WARP~\cite{Brunetti:2005hd} and Dark Side~\cite{Wright:2011vq}.

Dark Matter particles with masses in the \SIrange{10}{1000}{\giga\electronvolt} range or higher are expected to appear in the detector as nuclear recoils at energies up to roughly \SI{100}{\kilo\electronvolt}. The most prevalent backgrounds in this energy range are from the beta decay of \Nuc{Ar}{39} and from gamma and muon interactions. Electromagnetic backgrounds such as these can be suppressed by many orders of magnitude through PSD~\cite{Boulay:2004ut,Boulay:2009ug,PhysRevC.78.035801}.

The argon scintillation light has a wavelength of \SI{128}{nm}, which is not directly detectable with conventional photo-electron multiplier
tubes (PMTs). However, wavelength shifting dyes are available which efficiently shift the liquid argon scintillation light to the visible range.

As is common for many scintillators, nuclear recoils and electron recoils have different scintillation efficiencies in liquid argon. If the energy scale for nuclear-recoil events is denoted by the subscript ``R'' and the scale for electron-recoil events is denoted by the subscript ``ee" then the nuclear-recoil quenching factor, defined as $E_{R}/E_{ee}$, for liquid argon is about 0.25~\cite{Gastler:2012ki}.

\section{The DEAP-1 detector} \label{sect:deap1}
DEAP-1 was built as a
prototype for the DEAP-3600 detector~\cite{DEAPTaup2012} currently under construction at SNOLAB~\cite{Duncan:2010fq}. It was used to study the pulse shape
discrimination power in liquid argon, to study detector backgrounds,
and for prototyping components for the DEAP-3600 detector. Results from
the initial run at Queen's University were previously
reported~\cite{Boulay:2009ug}. Here, we report on the
background studies done with three iterations of the DEAP-1 detector
since the start of operation underground at SNOLAB. 

\begin{figure}[htbp]
   \centering
	\includegraphics[width=\textwidth]{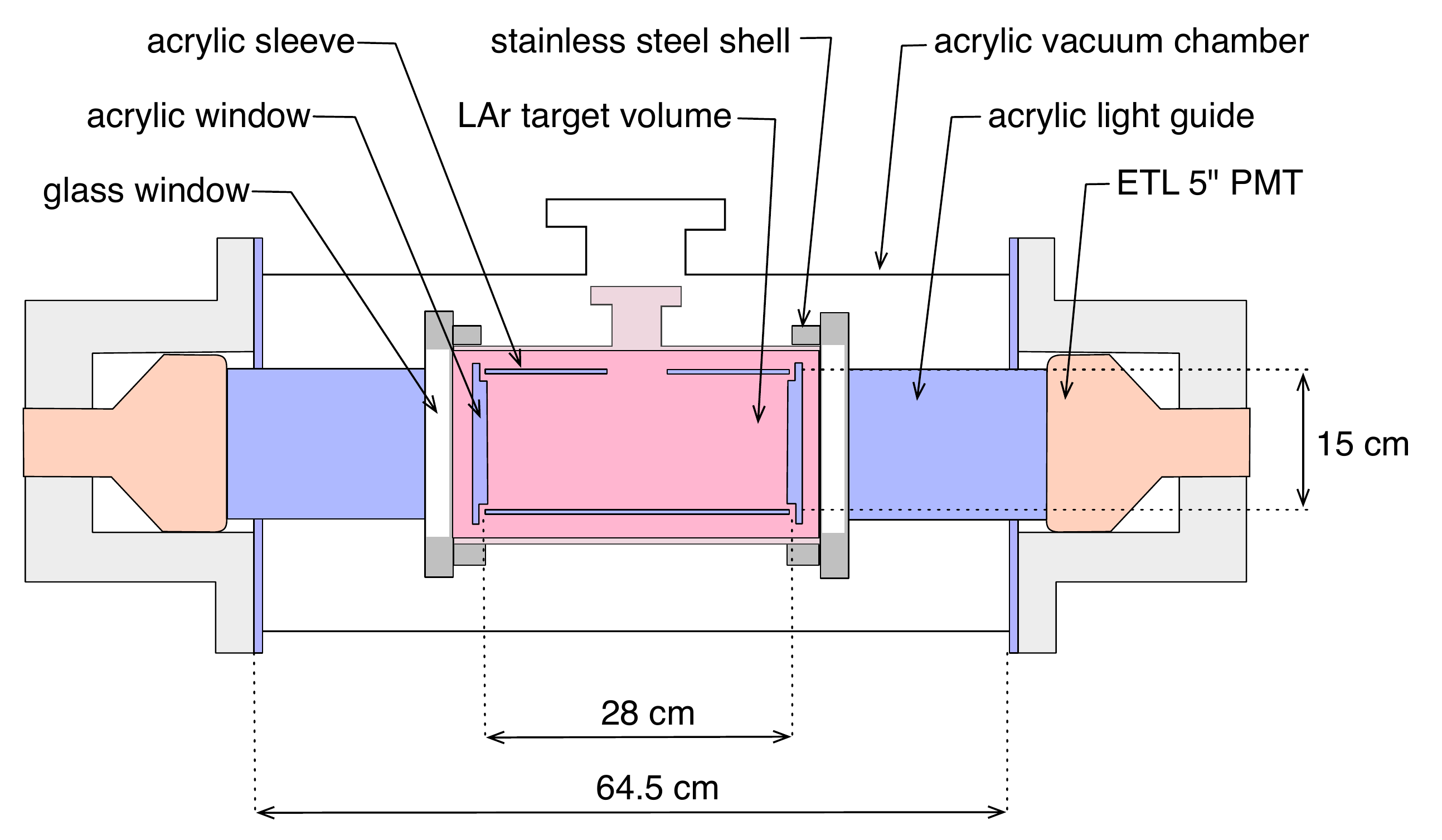}
   \caption{Schematic drawing of the DEAP-1 detector (G1). Different detector generations used different PMTs and different acrylic window configurations. The acrylic sleeve was wrapped in a diffuse reflector, which is not shown. }
   \label{fig:deap1}
\end{figure}

The DEAP-1 detector is schematically shown in Fig.~\ref{fig:deap1}.
The active volume is a cylinder, \SI{28}{\centi\metre} long
with a \SI{15}{\centi\metre} diameter, defined by an acrylic sleeve \SI{0.64}{\centi\metre}
thick, and closed off on each end with acrylic windows. The sleeve and acrylic windows are coated on the inside by vacuum deposition with the wavelength shifter
1,1,4,4-tetraphenyl-1,3-butadiene (TPB), which shifts the liquid argon
scintillation light to a peak wavelength of
440~nm~\cite{Davies:1996vd}, where acrylic is transparent and PMTs are sensitive. The TPB layers varied in thickness between the detector generations, and between the windows and the acrylic sleeve, in a range from \SIrange{1}{4}{\micro\metre}. 

The outside of the sleeve is covered by a teflon
and Gore diffuse reflector and contained in a cylindrical stainless steel
vessel with glass windows at each end. The total mass of liquid argon
inside the vessel is \SI{7.6}{kg}, with \SI{7.0}{kg} in the active volume. Acrylic light guides with PMTs at their ends are attached to the glass windows. The
light guides serve the dual purpose of moderating neutrons from the
PMT glass and thermally insulating the PMTs from the liquid argon
volume, thus allowing them to be operated at near room temperature. 

An opening at the top of the acrylic and stainless steel cylinders
leads to the ``neck'', a pipe through which liquefied argon is filled
into the detector.  During normal operation argon gas flows
up the neck, through a condenser, and back to the detector volume. The argon is purified only during the initial fill, but is continuously recirculated for cooling.

DEAP-1 went through three major iterations since the start of
operations at SNOLAB. In the first generation, G1, 5~inch ETL~9390B
(flat-face) PMTs were used.  The acrylic windows loosely fit the
acrylic sleeve, and the neck opening was \SI{2.54}{\centi\metre} in diameter.  

In the second generation, G2, the PMTs were replaced by Hamamatsu
R5912 8~inch high quantum efficiency PMTs, which will be used in
DEAP-3600. The light guides were replaced to accommodate the larger
and differently shaped PMTs.  

The third generation, G3, had a re-designed acrylic sleeve with a \SI{0.64}{\centi\metre} radius neck opening covered with a plug that let liquid argon in
but prevented light generated in the neck from entering the active
volume. The G3 windows were machined to fit tightly against the acrylic
sleeve, again to prevent gaps through which scintillation light could enter the target
volume. 

Data in G2 and G3 were taken with both nominal (NV) and lower than nominal voltage (LV) on the PMTs to increase the dynamic range. NV data has good energy resolution at energies below 500~keV, while LV data has good energy resolution at energies above 4.5~MeV. The lifetimes quoted in the figures, and all analysis at energies below \SI{4}{\mega\electronvolt} are for the NV data sets. The LV data sets were used only for study of the high energy alpha spectra.

\begin{table}[htbp]
   \centering
   \begin{tabular}{@{} lllll @{}} 
      \toprule
      Detector generation    & Start date & End date & \multicolumn{2}{c}{Livetime [d]} \\
        & & & NV & LV \\
      \midrule
      G1      &  07/2009 & 12/2009 & 34.8 & -  \\
      G2       &  03/2010 & 09/2010  & 37.8 & 14.0 \\
      G3       &   06/2011 & 04/2012 & 26.7 & 13.4 \\
      \bottomrule
   \end{tabular}
   \caption{The start and end dates of operation for the three detector generations, and the total lifetime of the data runs used for analysis.}
   \label{tab:booktabs}
\end{table}

\section{Background contributions to DEAP-1} \label{sect:d1background}

Liquid argon is sensitive to ionizing radiation, including particles
from cosmic rays, external and internal
gammas and electrons, alpha particles, and neutrons. While all of these radiation types cause signals in the detector, nuclear-recoil-like signals at energies below \SI{120}{\kilo\electronvolt} recoil energy directly compete with a possible WIMP signal. The other types of radiation cause background events only when misidentified as nuclear-recoil-like signals (PSD leakage).

External electromagnetic background from muons is reduced to \SI{0.29}{m^{-2}.day^{-1}}, by locating the
DEAP-1 detector underground at SNOLAB at a depth of 6010 meters water
equivalent~\cite{snolabhandbook}.

Alpha particles exciting liquid argon produce a signal similar to nuclear recoils and are thus a source of background not discriminated by the
PSD.  We expect alpha particles from alpha emitting isotopes in the
primordial \Nuc{U}{238} and \Nuc{Th}{232} decay chains to interact in
the DEAP-1 and DEAP-3600 detectors through those isotopes' presence
in detector and process system materials.

\begin{figure}[htbp] 
   \centering
	\includegraphics[width=0.9\textwidth]{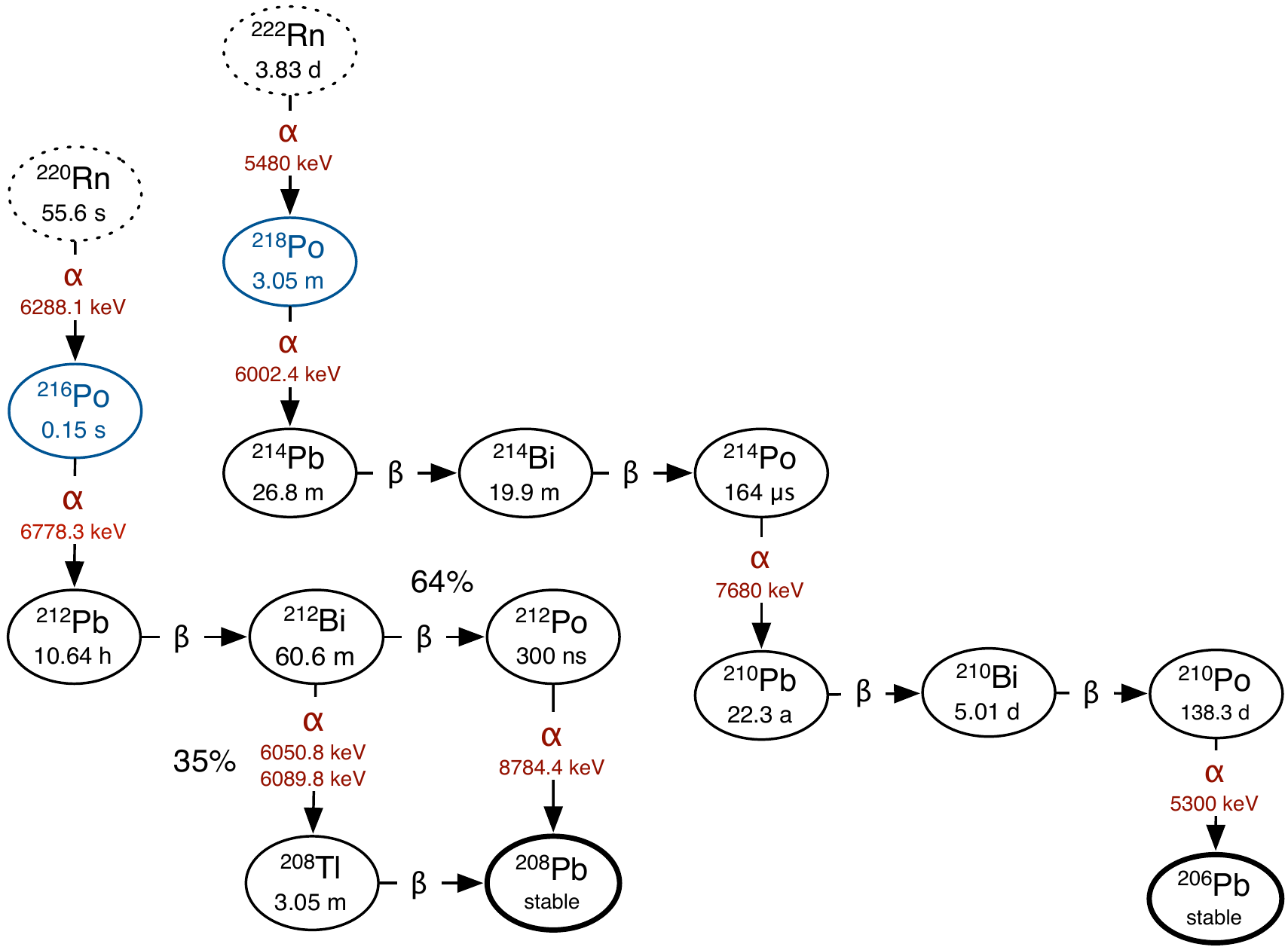}   
\caption{Decay chains of $^{220}$Rn (from \Nuc{Th}{232}) and $^{222}$Rn (from \Nuc{U}{238}) showing the relevant alpha decays for DEAP-1 and DEAP-3600. }
   \label{fig:rn220}
\end{figure}

Both chains, partially shown in Fig.~\ref{fig:rn220}, contain radon
isotopes which can emanate into the liquid argon and circulate through
the process system. It takes about 20~minutes for argon gas just
above the chamber to circulate through the argon return and cooling
lines.  The \Nuc{Rn}{220} from the thorium chain has a half-life of
only 56~seconds and thus must emanate close to the detector chamber to
produce background events. The \Nuc{Rn}{222} from the uranium chain, with a
half-life of 3.8~days, can reach the detector chamber from any
emanation location in the system. Thus uranium-chain backgrounds are expected to
dominate over thorium-chain backgrounds. 

The alpha particles from primordial isotope chains all have
energies in excess of \SI{5}{MeV}, which is significantly above the WIMP region of interest. Full energy alpha events, located in the active
liquid argon volume where scintillation light is detected at high
efficiency, can thus be readily discriminated based on their energy.

There are two scenarios in which the visible energy of an alpha
particle is reduced, allowing it to appear in the energy region of
interest for WIMP search. First, the event can occur at a location where
the efficiency for collecting light is low, such as in the gap between the
acrylic window and the acrylic sleeve or in the neck of the
detector. Second, isotopes embedded in or
adhering to the surface of the acrylic window or sleeve can alpha decay, resulting in reduced
visible energy because the alpha particle loses energy in the acrylic
or TPB layer before entering the liquid argon. In the second case, a recoil nucleus can be emitted into the argon while the accompanying alpha particle travels into the surface, causing a low energy nuclear recoil event.

Neutrons can reach the detector's active volume and induce a WIMP-like event.
Because of the highly suppressed muon rate at SNOLAB, muon spallation-induced neutrons 
are not a significant background source and neutrons produced in 
nuclear fission and ($\alpha$, n) reactions, induced by alphas from the $^{238}$U and $^{232}$Th chains, 
become dominant.

Samples of materials used in the construction of DEAP-1 were assayed
for U/Th content in the SNOLAB germanium counter~\cite{SnolabGE}. A GEANT4~\cite{Geant4} based Monte Carlo simulation was
used to estimate the expected number of neutrons produced from these radioactive
impurities. The rock wall neutron rate at SNOLAB was taken from Ref.~\cite{Heaton}.
Neutron production in the aluminum dark-box, stainless steel stand for the water shield (see Sect.~\ref{shielding}), and rock wall is expected to lead to a rate of about 10 n/year/kg in the region of interest, for a pulse shape discrimination parameter cut corresponding to 85\% detection efficiency for nuclear recoils.  The energy region of interest is \SIrange{30}{50}{keV_{ee}} and the event position region of interest is a volume given by \SI{+5}{\centi\metre} to \SI{-5}{\centi\metre} from the centre of the detector along its axis with a radius equal to that of the acrylic sleeve. We calculated that in the same region of interest, the inner detector (steel chamber and PMTs) contributes less than \SI{0.1}{n/year/kg}.

\section{Background reduction measures} \label{sect:bgreduction}
\subsection{Chamber preparation}
The inner acrylic chamber consists of the acrylic sleeve and the
windows. The goal of a careful chamber preparation was to minimize the
contamination of the detector surfaces with alpha emitting radon
daughters.

The acrylic sleeve of DEAP-1 was machined out of commercially
available acrylic stock from United Resins, Inc. The acrylic for the
windows was acquired from Spartech Townsend Plastics. The radon diffusion length in acrylic is
\SI{0.17}{mm}~\cite{1991NIMPB..61....8W}, so a few
millimetres of acrylic were machined off each surface of the
chamber to remove radon daughters diffused into the
material. 

After machining, the chamber was exposed to air in a class 2000 clean room for approximately two hours, where it was washed in an ultrasonic bath with detergent solution and then with ultra-pure water. The air activity in the cleanroom
was measured to be \SI{10}{\becquerel\per\cubic\metre} of $^{222}$Rn with a commercial RAD~7 radon detector.  Exposure to air was then minimized through the use of a nitrogen-purged glove box with an overpressure equivalent to \SI{2.5}{\centi\metre} of water. The glove box was
constructed from materials with low radon emanation (acrylic, copper) and
has radon impermeable seals (teflon, butyl rubber). 

In order to remove the surface deposits and sub-surface implantation of \Nuc{Po}{210} accumulated after machining, \SI{20}{\micro\metre} of the acrylic windows and sleeve of G1 and G3 were sanded off with sandpaper chosen for low U and Th contamination. The procedure was tried on acrylic test plates with scratches of calibrated depth and it was determined that manual wet sanding would remove the required amount of material. The pieces were washed with pressurized ultra-pure water to remove the sanding debris and sandpaper residue. The
pressure wash was repeated 3 times, each time followed by drying (all steps in the purged glove box).

The chamber was subsequently exposed to cleanroom air for $\sim$1~minute while moving it from the glove
box to the TPB evaporator and back

The acrylic sleeve and windows in the second iteration (G2) were
coated on the inside with $\sim$\SI{80}{\micro\metre} of purified acrylic. Close
to 20 grams of clean acrylic were obtained by UV-induced
polymerization of vacuum distilled methyl methacrylate, the acrylic
monomer. The clean material was then dissolved in a solvent,
acetonitrile, which had been purified by vacuum distillation, and applied
to the inner chamber surfaces. 
The coating was cast on the acrylic window, and spin-coated on the cylindrical sleeve, then left to dry
(for details see Ref.~\cite{2011AIPC.1338..101K}).

In all three cases, TPB coatings were evaporated on the inner surface
of the chamber using the apparatus described in \cite{Pollmann:2011vf}.  The coatings on the acrylic windows were
deposited from a TPB-filled quartz crucible heated with a nickel-chromium
wire. The coating on the acrylic sleeve was deposited from a TPB
filled NiCr mesh installed along the cylinder axis. The crucible, NiCr
mesh and heater were washed beforehand in 10\% nitric acid followed by
ultra-pure water, in order to remove any traces of surface radon daughter
activity.

For the G2 and G3 coatings, the TPB was pre-heated to \SI{150}{\celsius} for a few
hours to evaporate potentially present contaminants with a lower evaporation temperature
than TPB (such as volatile organo-metallic compounds of polonium,
see~\cite{Mabuchi:1963wk}) in absence of the substrate. The
coating deposition rate and thickness during the final process was
monitored by quartz balance deposition monitors. 

The coated parts were stored in the dark in order to avoid TPB
degradation reported in some literature sources~\cite{Chiu:2012ur}.

Finally, the inner chamber was installed in the ultrasonically cleaned DEAP-1 outer
stainless steel vacuum chamber, which was then sealed, pumped out and
over-pressurized with 18~psi of nitrogen gas to minimize exposure to radon during transportation
between Queen's University and the underground site at SNOLAB.
Connecting the detector to the DEAP-1 process systems required a
$\sim$30 minute exposure to the ambient air with a \Nuc{Rn}{222} activity of about \SI{120}{Bq.m^{-3}}~\cite{snolabhandbook}.

\subsection{Argon purity and radon trap}
 
The argon gas for the DEAP-1 detector was taken from pressurized bottles of grade 5 argon from Air Liquide, and was further purified with a commercial argon purifier (SAES getter) resulting in ppb-level contamination from volatile hydrocarbons, water vapour, carbon dioxide, oxygen and nitrogen. The low level of radon which emanates from the gas bottles or the attached armature was removed with a dedicated radon trap.
 A tube filled with \SI{10}{g} of activated carbon spheres
(Carboxen\textsuperscript{\textregistered}) was placed on the argon inlet between the argon source
and the liquefying column. The trap was cooled in a bath of ethanol and liquid nitrogen to \SI{-110}{\degreeCelsius} with no transient excursions warmer than
\SI{-100}{\degreeCelsius}. Tests of the trap indicated that less than one atom of $^{222}$Rn
passed the trap and enterd the detector during a fill~\cite{Anonymous:d9TENHx4}.

\subsection{Shielding} \label{shielding}
The detector is shielded against neutrons by a
minimum of two layers of ultra-pure water in \SI{30.5}{\centi\metre} cubical
polyethylene containers, for a total water shielding thickness of \SI{60}{\centi\metre}, surrounding the entire detector.
External gammas and muons, at the level present at SNOLAB, are not a major concern for DEAP-1, so no additional gamma-shielding was used.

\subsection{Pulse shape discrimination}
The time spectrum of the emitted argon scintillation light depends on
the linear energy transfer (LET) along the track of the exciting
radiation~\cite{Hitachi:1983ja}. Pulse shape discrimination thus allows
separation of electron recoil interactions and nuclear recoil
interactions in the argon which have low and high LET
respectively. 

\section{Data analysis} \label{sect:dataanalysis}
Data from the detector is stored in the form of digitized PMT traces, one \SI{16}{\micro\second} trace from each PMT per event. The traces are analyzed off-line to find, among other parameters, the total number of photo electrons (TotalPE) for the event, the pulse shape parameter Fprompt, and the location of the event along the axis of the detector (Zfit).

TotalPE is the total charge of the event (integrated over \SI{10}{\micro\second}), divided by the charge created by a single photo electron. TotalPE is related to event energy by the light yield, which is calibrated using \Nuc{Na}{22}
(\SI{511}{keV}) and \Nuc{Am}{241} (\SI{59.5}{keV}) gamma
calibration sources. 
At the \SI{59.5}{keV} peak, the light yields are (2.4 $\pm$0.2)~PE/keV, (4.7$\pm$0.5)~PE/keV and (4.0$\pm$0.4)~PE/keV for G1, G2 and G3 respectively. The light yield changes in a non-linear way when going to higher energies. The uncertainties are due to the uncertainty of the single photo electron (SPE) charge and fluctuations in SPE charge over time. The difference in light yield between G2 and G3, which have the same PMTs, can be explained by differences in TPB thickness and by an improved method in G3 of determining SPE charges using an external light source.

Fprompt is the charge accumulated, or the fraction of light emitted, within the first \SI{150}{ns} of an event (the prompt charge), divided by the total integrated charge of the event. The values the Fprompt parameter can take are logically bounded by $[0, 1]$.

\begin{figure}[htbp] 
   \centering
   \includegraphics[width=\textwidth]{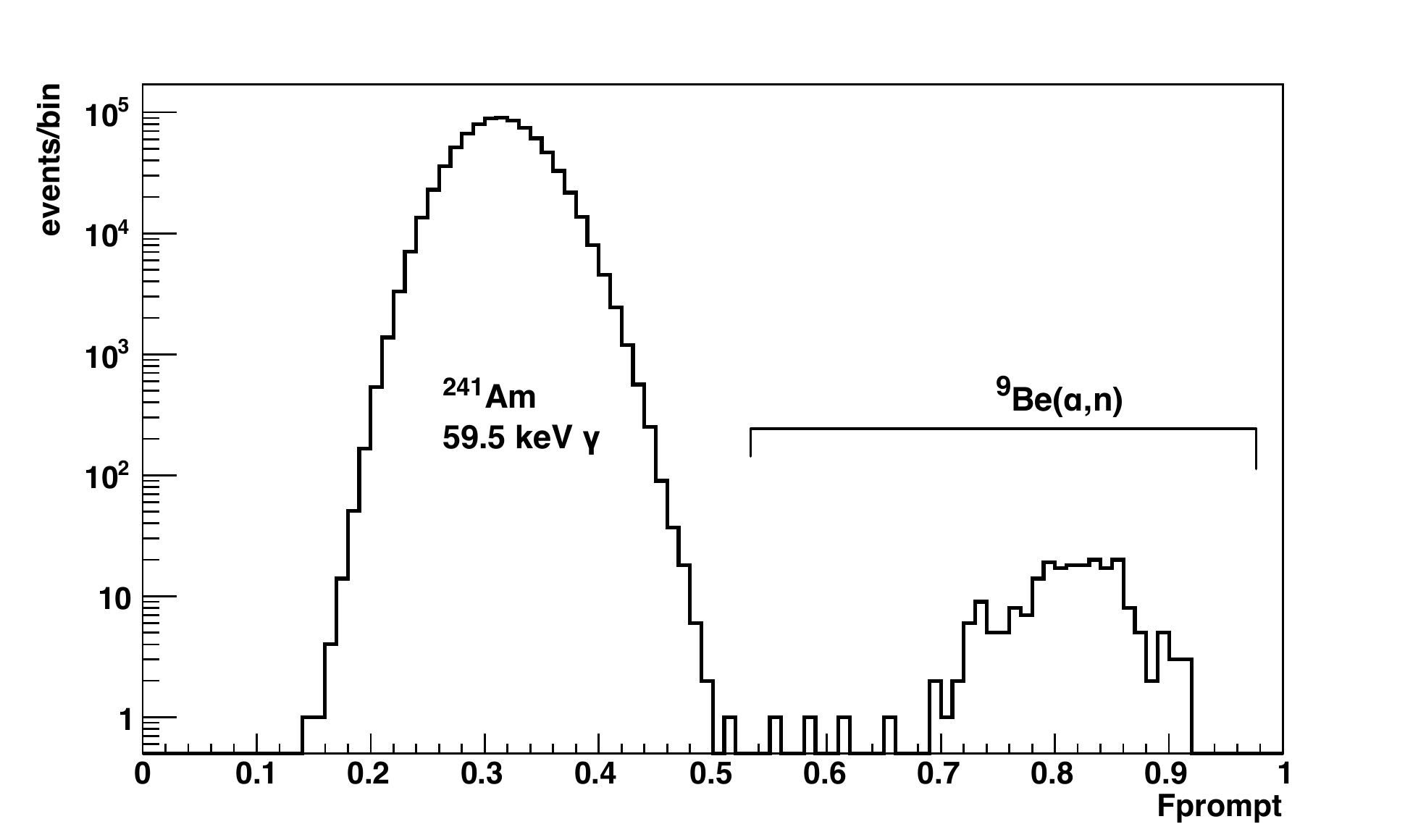} 
   \caption{Fprompt distribution of gammas and neutrons from an AmBe calibration run in DEAP-1 G3, for events at the \SI{59.5}{keV_{ee}} gamma peak and with standard cuts. The events between 0.5 and 0.7 Fprompt are due neutrons from the source. }
   \label{fig:V5FpDistribution}
\end{figure}

Nuclear-recoil events are identified and separated from electron-recoil events solely by the Fprompt parameter. 
The shape of the Fprompt distributions at different energies is determined using a tagged \Nuc{Na}{22} source (as described in Ref.~\cite{Boulay:2009ug}) for electron recoils and an AmBe neutron source for nuclear recoils. Electron-recoil interactions have a mean Fprompt value around 0.3, compared to 0.7 to 0.8 (depending on the PMTs and DAQ electronics and on the event energy) for nuclear recoil interactions, as shown in Fig.~\ref{fig:V5FpDistribution}. For the purpose of choosing cuts, the shape of the neutron Fprompt distribution can be adequately modelled by a gaussian distribution.

Events with Fprompt in $[F_c, 1]$, where $F_c$ is a value of Fprompt between the mean for electron recoils and the mean for nuclear recoils, are called nuclear-recoil-like. Since electron-recoil events have a mean Fprompt lower than nuclear-recoil events, this cut can eliminate most of them from the data set. However, because the two distributions overlap, a certain number of electron-recoil events is still accepted by the cut, and some number of nuclear-recoil events are lost. We choose $F_c$ so that it minimizes the gamma leakage while maximizing nuclear-recoil efficiency. A nuclear-recoil efficiency of 85\%, for example, means $F_c$ is chosen to retain 85\% of nuclear-recoil events. 

The ideal value of $F_c$ is dependent on energy since, for a fixed $F_c$, the gamma leakage increases towards lower energies. We typically choose a cut corresponding to 85\% nuclear-recoil efficiency to get the best possible statistics when studying nuclear-recoil-like background events. A nuclear-recoil cut of 50\% reduces the number of nuclear-recoil events in the data set, but also allows analysis at lower energies by removing more gamma-leakage events.

Zfit is determined from the relative amount of light detected in the two PMTs and indicates the position of events along the axis of the detector joining the two PMTs. The centre of the detector is then at \SI{0}{\centi\metre}. The Zfit parameter was
calibrated using the back-to-back \SI{511}{keV} gamma rays from an external
\Nuc{Na}{22} source. A tag NaI detector was held behind the source and
tuned to trigger when a backward-going \SI{511}{keV} gamma was measured. The
source and tag detector were scanned along the length of the
DEAP-1 active volume, allowing readout to be triggered only on gamma events at a
specific z-location within DEAP-1.
The detector has no spatial resolution in the x and y-directions.

All data shown here have a basic set of hardware cuts applied to them, to remove PMT traces where no pulse was found, where there was more than one pulse (pile-up event), where the pulses from the two PMTs did not start at the same time, or where the traces obviously did not form a proper scintillation pulse shape, namely where the charges in the prompt or late integration windows were positive. 

\begin{figure}[htbp] 
   \centering
	\includegraphics[width=0.9\textwidth]{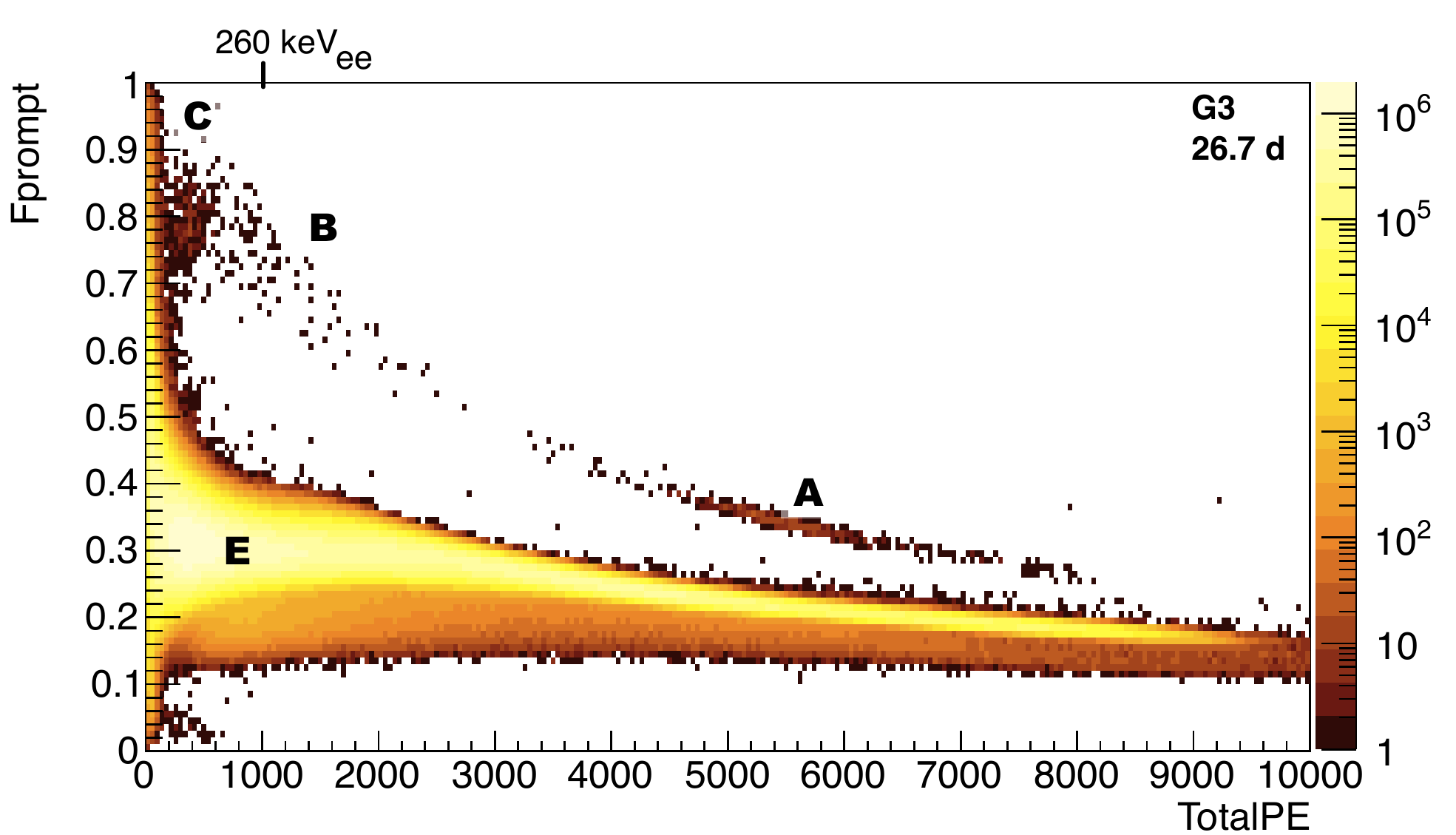} 
	\caption{G3 background data with basic cuts. The event populations identified by A, B and C correspond to those in Fig.~\ref{fig:zfitvstotalpe}. E denotes the electron-recoil band, which is dominated by betas from \Nuc{Ar}{39}.}
   \label{fig:fpvstotalpe}
\end{figure}

The G3 background data with only the basic cuts applied is shown in Fig.~\ref{fig:fpvstotalpe}. The event populations A, B and C are in the nuclear-recoil band. The band does not remain at a constant value of Fprompt when going to higher energies because of saturation effects in the photon readout chain. The population identified as E is from electron recoils and forms the electron-recoil band. The saturation affects nuclear-recoil and electron-recoil pulses differently because of the difference in the intensity of the prompt peaks.

Cuts on Fprompt, TotalPE and Zfit are chosen based on the event populations under consideration. Their correspondence to nuclear recoil efficiency, energy and location in the detector was calibrated for each detector generation, and changes with energy due to saturation effects.
The low energy region-of-interest goes up to about \SI{260}{keV_{ee}}, with $F_c$ chosen to be either 85\% or 50\% nuclear recoil efficiency. The high energy region-of-interest is above \SI{4.5}{MeV_r} (approximately 4500~TotalPE in G3). The PMT response is non-linear at high energies, so a sloped cut in $F_c$ is chosen that follows the upper edge of the electron-recoil band. There is complete separation between the bands at high energies, so the Fprompt cut there has 100\% nuclear recoil efficiency and approximately 0\% gamma leakage.

\section{Background analysis} \label{sect:bganalysis}

\begin{figure}[htbp] 
   \centering
	\includegraphics[width=\textwidth]{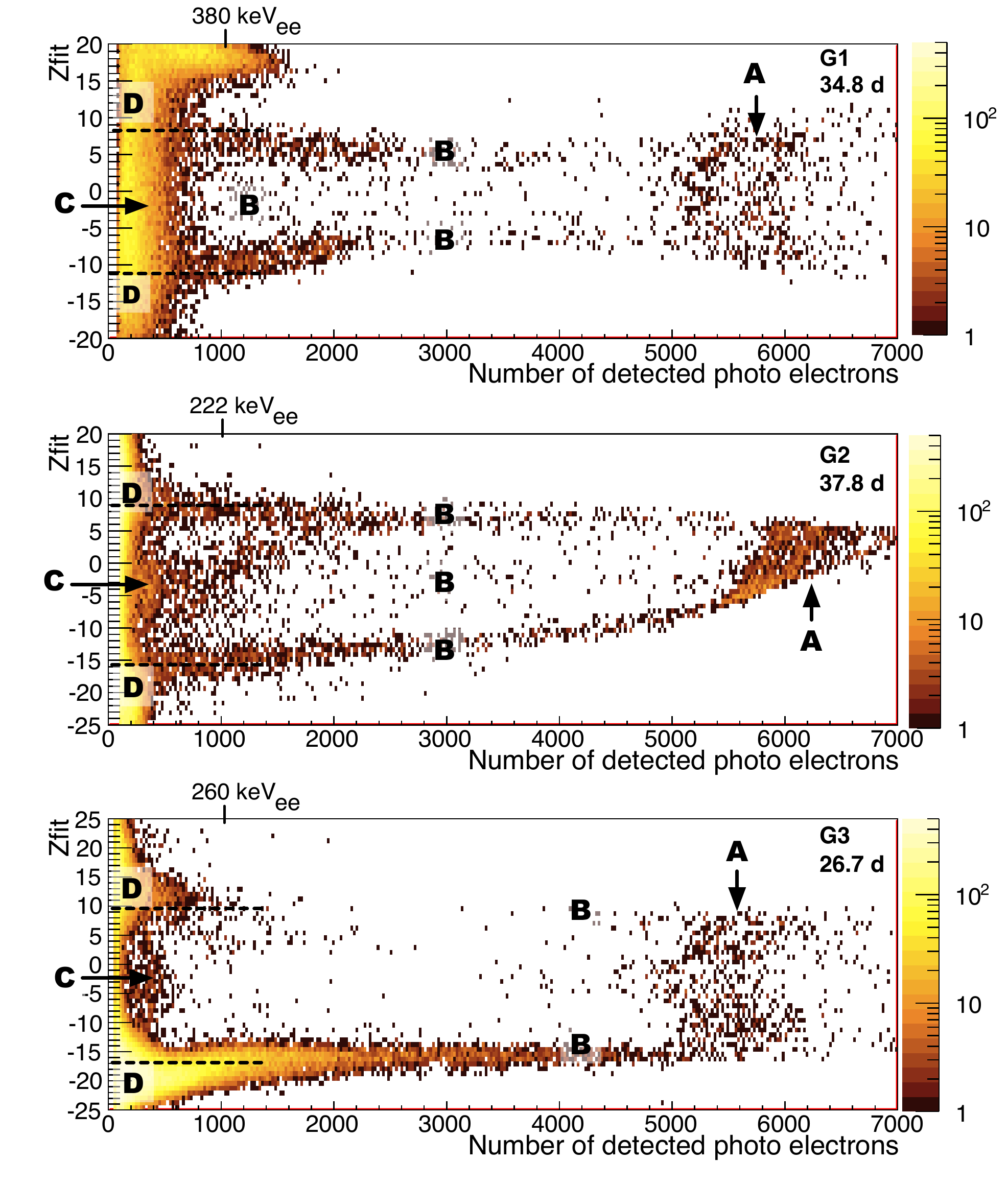}   
	\caption{The reconstructed location of nuclear-recoil-like events along the axis
          of the detector (Zfit approximately in units of cm) versus the number of detected photo electrons. The three panels correspond to DEAP-1 detector generations G1 through G3, with respective live-times quoted for each. The number of photo electrons corresponds to energies between 0 and 8~MeV. Dashed lines indicated the approximate location of the acrylic windows for low energy events. Saturation effects above about 2000~PE cause the Zfit calibration to be nonlinear. Different saturation points of the two PMTs cause the uneven distribution of population~A in G2. The curved shape of population A in G1 and G3 is due to a lower light collection efficiency near the centre of the detector.}
   \label{fig:zfitvstotalpe}
\end{figure}

Fig.~\ref{fig:zfitvstotalpe} shows the location of nuclear-recoil-like events in the DEAP-1 detector versus the number of detected
photo-electrons for the three detector generations. Four distinct populations of events are visible in each generation: A high
energy population evenly distributed across the detector (A), a
population ranging in energy from zero up to the high energy
population but located mainly near the windows and in the centre of
the detector (B), a population at low energies evenly distributed
across the detector (C), and a population near the windows of the
detector at low energies (D). The origin of each population will be
discussed in the following sections. A more detailed discussion can be found in Ref.~\cite{pollmann}.

\subsection{Analysis methods}
\subsubsection{Coincidence tagging radon chain sections} \label{sec:tagging}
Timing coincidence tags and analysis of the energy
spectra indicate that population A in Fig.~\ref{fig:zfitvstotalpe} is from decays of
radon and its daughters in the active liquid argon volume.

The polonium daughters of $^{220}$Rn and $^{222}$Rn have short half-lives, 0.15~s and
183~s respectively. A pair of consecutive events in population A is preliminarily tagged as a $^{220}$Rn coincidence if the two events are separated by up to $\Delta t= 0.5$~s, and as a $^{222}$Rn coincidence if the pair is separated by 0.5~s to $\Delta t = 500$~s. 

To find the number of real coincidences, $N_{c}$, the expected number
of random coincidences $N_{r}$ must be subtracted from the number of
tagged events $N_{tag}$ and the efficiency of the tag must be taken
into account. The number of random coincidences can easily be
estimated if the total number of events in population A, $N_{all}$, is much larger than
$N_{c}$. This is not the case in much of the DEAP data, especially
during the radon calibration spike. The actual number of radon coincidences must
thus be calculated as
\begin{align}
N_{c} & = [N_{tag} - N_{r}]/\epsilon \\
	& = [N_{tag} - (N_{all}- N_{c}) \cdot (1 - e^{-(N_{all}-N_{c})\cdot\Delta t/T} )]/\epsilon
\end{align}
where T is the live-time of the data taking run and $\epsilon$ is the efficiency of the tag. For the time windows used above, the efficiencies are 99.9\% and 85\% for the $^{220}$Rn and $^{222}$Rn tags.
This equation has no analytic solution, however an iterative solution can be found where
\begin{align}
N_{c}^{(0)}& = 0 \\
  N_{c}^{(n)}& =[N_{tag} - (N_{all} -N_{c}^{(n-1)})  \cdot (1 - e^{-(N_{all} -N_{c}^{(n-1)})\cdot\Delta t/T} )]/\epsilon
\end{align}
This converges after 6 to 11 iterations in our data sets. 

\begin{figure}[htbp] 
   \centering
\includegraphics[width=1.0\textwidth]{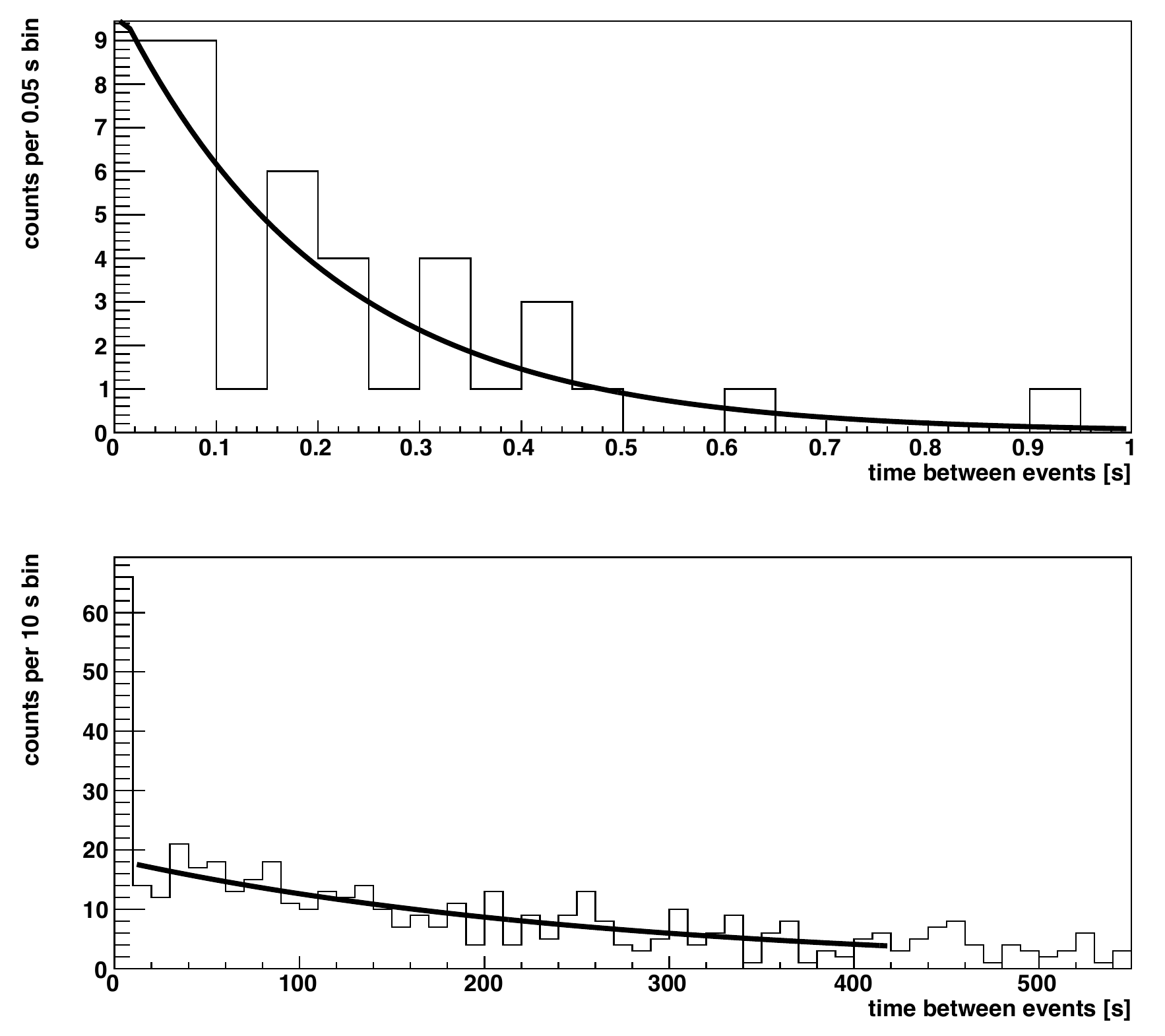}
\caption{The timing spectrum for events from population A (G2). Top: Up to 1 second, fitting a half-life of 0.14$\pm$\SI{0.03}{s} consistent with the \Nuc{Po}{216} decay. Bottom: Up to 420 seconds, fitting a half-life of 181$\pm$\SI{26}{s}, consistent with the \Nuc{Po}{218} decay.}
   \label{fig:timing}
\end{figure}

The timing spectrum of pairs of consecutive events from population A is shown in Fig.~\ref{fig:timing}. The fitted half-lives are those of $^{216}$Po and $^{218}$Po, indicating that these tags are indeed radon coincidences.

\begin{figure}[htbp] 
   \centering
	\includegraphics[width=\textwidth]{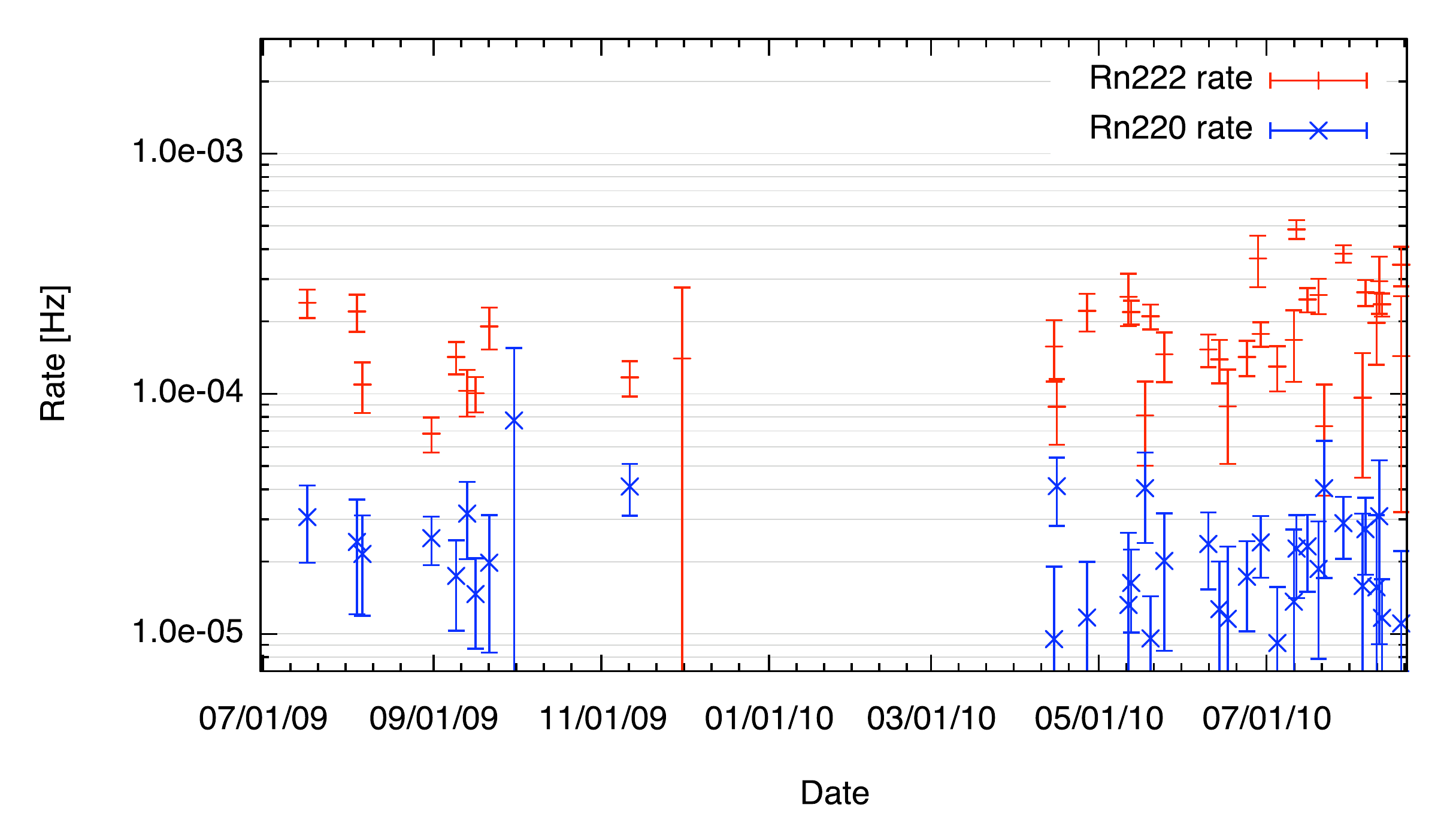}   
\caption{Tagged rates of $^{222}$Rn and $^{220}$Rn for each background
  run in G1 (data from 2009) and G2 (data from 2010) are roughly constant in time and consistent between detector generations.}
   \label{fig:taggedrates}
\end{figure}

The average radon rates found for each detector configuration are
presented in Table~\ref{tab:taggedradon}. The rates for each
background run in G1 and G2 are shown in Fig.~\ref{fig:taggedrates}.
The rates were stable over many months. 
\begin{table}[htbp]
   \centering
   \begin{tabular}{@{} llllll @{}} 
      \toprule
        Data set &  \multicolumn{2}{c}{Number of tags} &  \multicolumn{2}{c}{Rate [$10^{-6}$Bq] } \\  
       &  $^{222}$Rn  & $^{220}$Rn  &$^{222}$Rn & $^{220}$Rn\\
      \midrule
       G1 NV & 349 $\pm$22 & 77 $\pm$7 & 116 $\pm$7 & 26 $\pm$2 \\
	G2 LV & 199 $\pm$17 & 27 $\pm$5 & 163 $\pm$14 & 22 $\pm$4 \\ 
	G2 NV & 599 $\pm$29 & 62 $\pm$8 & 183 $\pm$9 & 19 $\pm$2 \\ 
	G3 LV & 135 $\pm$13 & 16 $\pm$4 &  116 $\pm$35 & 15 $\pm$3 \\
	G3 NV & 193	$\pm$17 & 29 $\pm$5 & 84 $\pm$7 & 13 $\pm$2 \\
      \bottomrule
   \end{tabular}
   \caption[Number of tagged radon decays in DEAP-1 G1, G2, and G3.]{Number of tagged radon decays in DEAP-1 G1, G2, and G3. LV and NV stand for lower than nominal voltage and nominal voltage runs respectively. To compensate for different lengths of data runs, the numbers are also given converted to rates.}
   \label{tab:taggedradon}
\end{table}

The so called ``Bi-Po'' tag, where one looks for coincidences between the beta particle from \Nuc{Bi}{214} and the alpha particle from \Nuc{Po}{214}, is commonly used in radon-chain analysis. 

The expected fraction of these coincidences relative to the total number of alpha events (the tagging efficiency) depends on the location of the decaying nucleus. If the decays happen in the liquid argon volume (not on a surface), it should be close to 100\%. If all of the \Nuc{Po}{214} is plated out on detector surfaces, the efficiency for detecting either the alpha or electron recoil depends on the event and surface topology. In the simplest case, half of alpha events should be preceded by betas. As for the other half, while betas enter the substrate and remain undetected, the de-excitation gammas, depending on their energy and emission angle, still have some chance of inducing detectable scintillation.  A GEANT4 based simulation of DEAP-1, coupled with the DECAY0~\cite{Tretyak} event generator, was used to estimate the Bi-Po tagging efficiency. It indicates that if \Nuc{Bi}{214} is plated out on the inner TPB surface, we expect that 64\%\ of \Nuc{Po}{214} events will be seen in coincidence with a preceding electromagnetic event in DEAP-1. A consistent result was also obtained with the native GEANT4 radioactive decay generator.

We observed Bi-Po coincidences for events in the high-energy \Nuc{Po}{214} peak (see Fig.~\ref{fig:V5HighE}). The rate of uncorrelated electron recoil events in the detector is such that we expect random coincidences for up to 10\%\ of the high-energy \Nuc{Po}{214} events. After subtraction of the random coincidence fraction we estimate the fraction of real Bi-Po coincidences as 65$\pm$10\% for G1 and 56$\pm$17\% for G3 (LV data), both consistent with the 64\% expectation.  

The above reasoning and the expected tagging efficiency estimate hold also for low-energy events caused by nuclear recoils from \Nuc{Po}{214} decays entering liquid argon and alphas, emitted simultaneously in the opposite direction, entering the substrate.  The low-energy \Nuc{Po}{214} signature of that kind is possible only when the decay happens on the detector surface, and again, if the \Nuc{Po}{214} population is fully plated out, in the simplest situation half of the decays will manifest as high- and the other half as low-energy events, with both event types preceded by \Nuc{Bi}{214} betas in 64\%\ cases.  Detailed analysis of Bi-Po tagging for low-energy events will be discussed in Sect.~\ref{sec:bipos}.

\subsubsection{Fitting full energy radon alpha spectra} \label{subs:rnspectra}

\begin{figure}[htbp] 
   \centering
\includegraphics[width=0.9\textwidth]{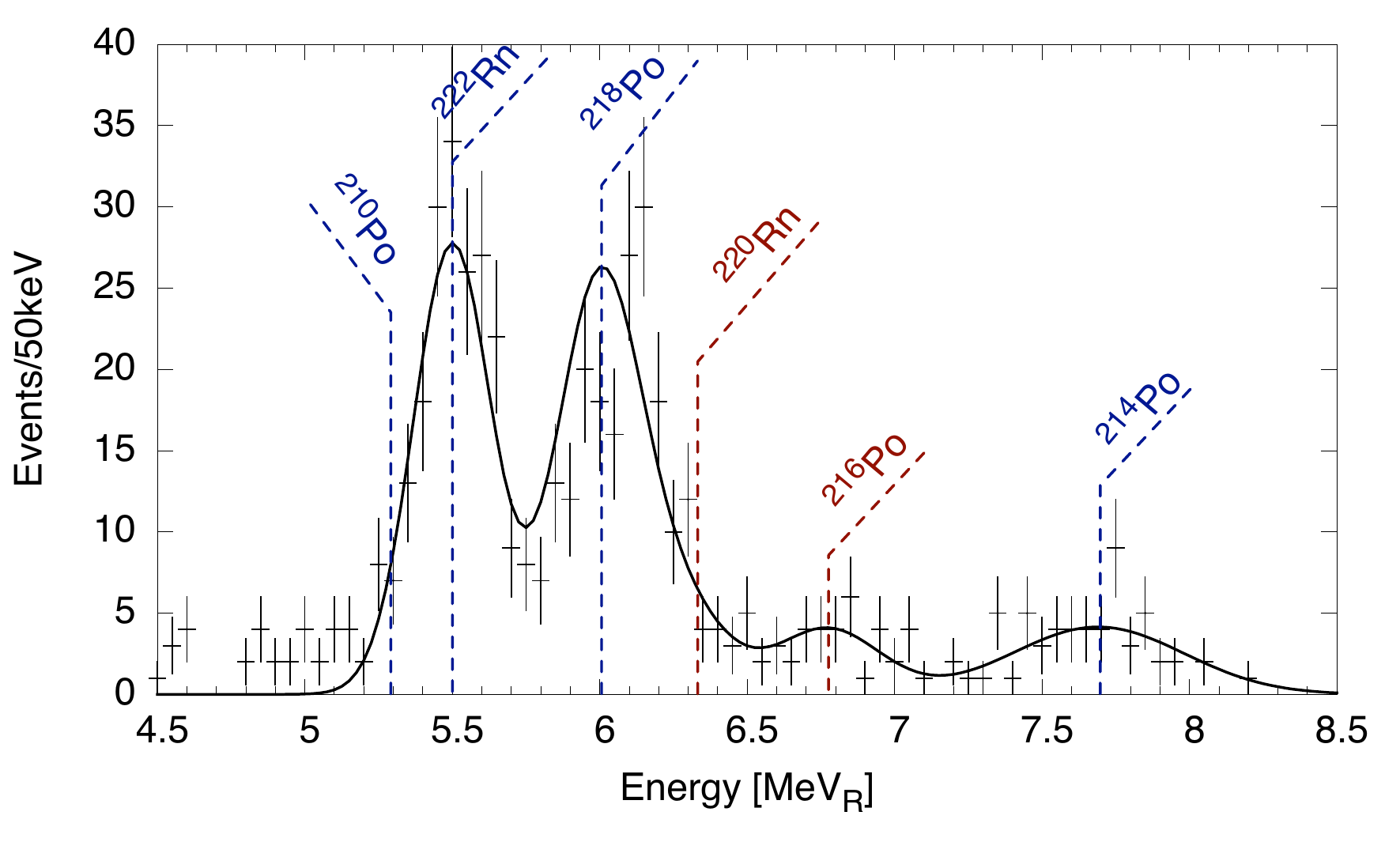}
   \caption{Spectrum of full energy alphas in DEAP-1 G2.}
   \label{fig:V3midasHighE}
\end{figure}

\begin{figure}[htbp]
   \centering
	\includegraphics[width=0.9\textwidth]{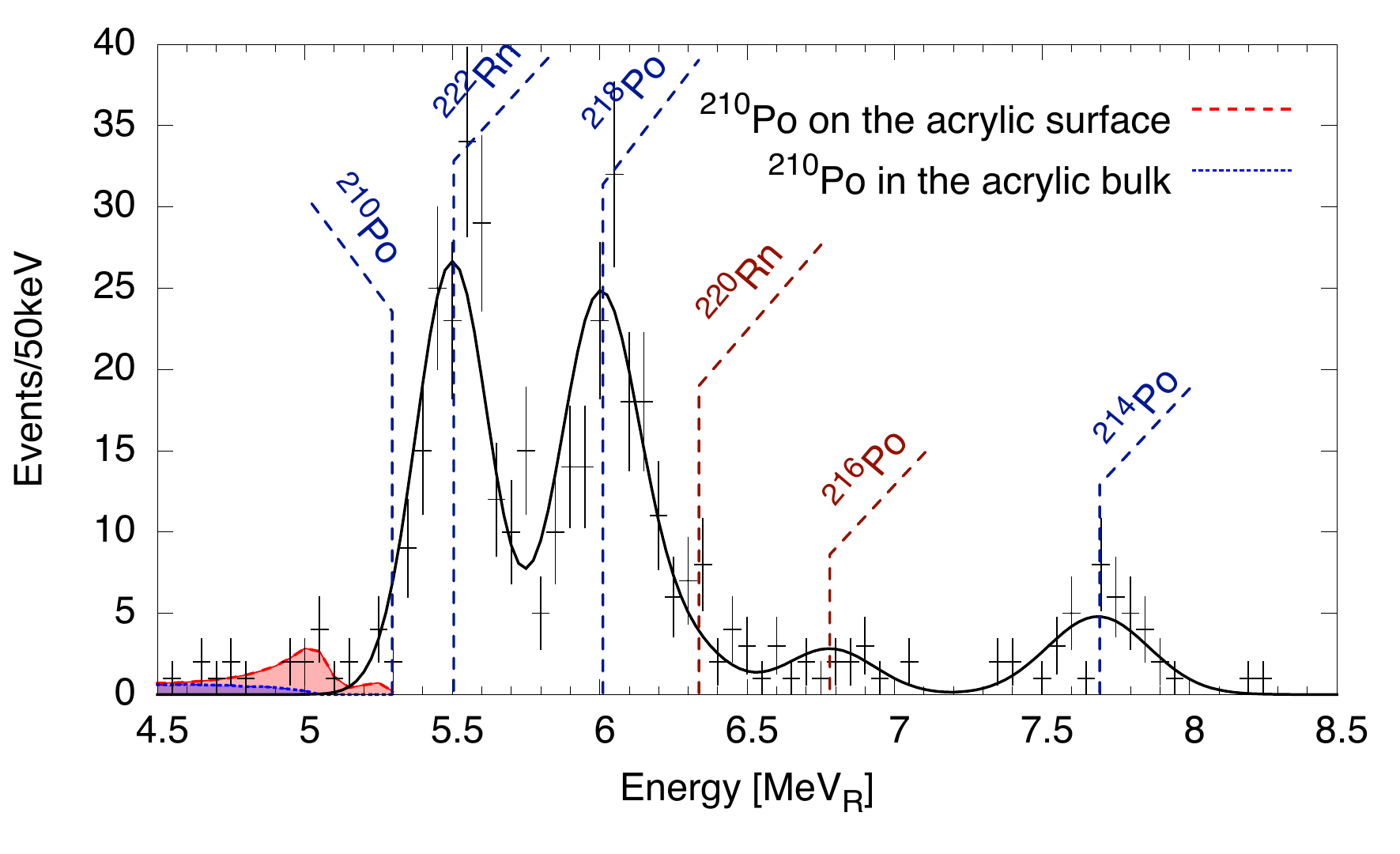}
   \caption{Spectrum of full energy alpha particles from the $^{222}$Rn and $^{220}$Rn chains in DEAP-1 G3. Also show is the simulated spectrum of $^{210}$Po on the acrylic surface, which was fitted to the observed spectrum between 4.5 and 5.2~MeV (the red shaded region). The spectrum of $^{210}$Po in the acrylic bulk is also drawn (the blue shaded region), with the normalization found from fitting an intermediate energy range (see Sect.~\ref{sect:surface}).}
   \label{fig:V5HighE}
\end{figure}

The full energy alpha spectra - that is spectra for events where the full event energy is visible - for DEAP-1 G2 and G3 low voltage (LV) data are shown in Fig.~\ref{fig:V3midasHighE} and
Fig.~\ref{fig:V5HighE}. Low voltage data is used here because of its good energy resolution and approximately linear PMT response. The energy scale was calibrated and corrected for event position using the tagged alpha events from radon and polonium, and the spectra were fit with a sum of five Gaussians,
which goes with the assumption that all events above 5.2~MeV are due to decays of
$^{222}$Rn, $^{218}$Po, \Nuc{Po}{214}, $^{220}$Rn and $^{216}$Po. The energies of the peaks and the relative intensity of each except the \Nuc{Po}{214} peak were fixed, while the number of events in each chain, energy resolution, and \Nuc{Po}{214} scaling factor were
floated in the fit. 

The necessity of the \Nuc{Po}{214} scaling factor, $\eta$, can be explained as follows: relatively long lifetimes of $^{214}$Pb and $^{214}$Bi (see Fig.~\ref{fig:rn220}) 
allow them to freely float around the detector and eventually adhere to a surface.
Then, in the case of a perfectly smooth surface exactly half of the decays of their daughter, \Nuc{Po}{214},
should emit the alpha particle into the surface, while the other half emit
it into the argon. If the surface material does not scintillate, the \Nuc{Po}{214} scaling factor would be exactly $2$. A surface that does scintillate, such as the TPB, but at a much lower scintillation efficiency than the liquid argon, still moves the event out of the full energy peak, so that the scaling factor would again be $2$, and there would be a corresponding population of events at lower apparent energy.

A larger scaling factor is possible when there are surfaces which the \Nuc{Po}{214} can adhere to but which are not visible from the active volume, so that a decay from there is either not detected at all or is only detected with degraded energy (see Sect.~\ref{susbs:geometric}). The ratio is also different for rough surfaces, such as the TPB surface on the inside of the acrylic vessel, where some of the particles emitted into the argon can travel a short distance and then enter TPB again. 

Decays from surfaces not visible to the PMTs, and decays on rough surfaces, can thus explain why we observe fewer than $1/2$ of the events expected.

The same effect, albeit smaller because of the shorter lifetime, can be expected for the \Nuc{Po}{218} peak. During the G2 radon spike (see Sect.~\ref{susbs:geometric}), where alpha spectra with good statistics were obtained, the area under the \Nuc{Po}{218} peak was in fact only 80-85\% of the area under the \Nuc{Rn}{222} peak. The fits of the lower statistics alpha spectra from background runs were more stable without an additional \Nuc{Po}{218} scaling factor though, and the difference this made for the fitted out \Nuc{Rn}{222} rate was smaller than the uncertainty of the fit.

As for the remaining alpha decays in the radon chains, the $^{212}$Bi and $^{212}$Po are expected to have negligible peak intensities in these spectra, because the rate of $^{220}$Rn is much lower than that of $^{222}$Rn, the
chain is split between $^{212}$Bi and $^{212}$Po, and their rate can be expected to be suppressed by at least the same amount as that of \Nuc{Po}{214}, as they adhere to the detector's surfaces as well. They are thus disregarded in the fit.
 The rate of $^{210}$Po is strongly
suppressed due to the long half-life of \Nuc{Pb}{210}, and there is no indication of additional \Nuc{Pb}{210} from out-of-equilibrium sources, so it is disregarded as well.


\begin{table}[htbp]
   \centering
   \begin{tabular}{@{} llllll @{}} 
      \toprule
	& $\chi^{2}$/NDF & \multicolumn{2}{c}{Number of decays} & Resolution & \Nuc{Po}{214} reduction \\
	&      &  $^{222}$Rn  & $^{220}$Rn &  $(\sigma$/E)& ($\eta$)\\
      \midrule
	G1 & 1.3 & 350 $\pm$15 & 73 $\pm$9 & 0.033 $\pm$0.002 & 3.4 $\pm$0.4 \\
	G2 LV & 1.1 & 181 $\pm$10 & 36 $\pm$6 & 0.025 $\pm$0.002 & 2.9 $\pm$0.4 \\
	G3 LV & 1.0 & 165 $\pm$ 9 & 21 $\pm$4 & 0.023 $\pm$0.001 & 4.0 $\pm$0.7 \\ 
      \bottomrule
   \end{tabular}
   \caption{Parameters obtained from fitting the full energy alpha spectra (compare Fig.~\ref{fig:V3midasHighE} and Fig.~\ref{fig:V5HighE}).}
   \label{tab:spectrafit}
\end{table}

The fit results are summarized in Table~\ref{tab:spectrafit}. The fitted number of radon events is consistent with the number of tagged events in Table~\ref{tab:taggedradon} for both G1 and G2 and for both radon isotopes. The tagged number of \Nuc{Rn}{222} events in G3 is lower and equals 82$\pm$9\%\ of the fitted value. We found evidence that one of the new acrylic windows was able to move slightly, opening and closing the gap between window and acrylic sleeve, so that argon could flow through the gap and radon daughters could drift in and out from behind the window, breaking the tag chains. See the end of Sect.~\ref{susbs:geometric}.

To check if any other sources of alpha particles could be contributing to the high energy events in G1 and G2, the total number of tagged events was compared to the number of high energy events. Each tagged event is followed by three more alpha events down the decay chain. The \Nuc{Po}{214} events however are suppressed by a factor of $\eta$, as fitted. We apply the same suppression factor to the decays following \Nuc{Pb}{212} in the \Nuc{Rn}{220} chain. The number of events identified by the tags is then
\begin{equation}
N_{id} = (2 + 1/\eta)\cdot N_{222Rn} + (2 + 1/\eta)\cdot N_{220Rn}
\end{equation}

We find that the numbers of events due to other backgrounds are 8$\pm$~60 out of 985 for G1, -14$\pm$~44 out of 516 for G2 LV and 17$\pm$~99 out of 2097 for all G2 runs. In all cases, the number of unidentified events is consistent with zero and allows for a high energy event rate not caused by radon and daughters of no more than \SI{3.6e-5}{Hz}. While the tagging is not 100\% efficient in G3, the spectrum above \SI{5.5}{MeV_{r}} shows no indication of significant contributions from sources other than radon.

\subsection{Background classes observed in the nuclear-recoil band}
\subsubsection{Geometric alpha backgrounds} \label{susbs:geometric}
Population B in Fig.~\ref{fig:zfitvstotalpe} arises from alpha decays of radon and its daughters in regions of the detector not completely visible to the PMTs. The geometry of the detector includes small gaps near the windows and near the neck from where only some of the scintillation light can reach the PMTs, so events from these regions have reduced visible energy. This explanation was suggested by GEANT4 Monte Carlo simulations and
verified by an injection of $^{222}$Rn into DEAP-1 G2.

The radon spike was prepared using radon from SNOLAB air, which has an
activity of approximately \SI{120}{\becquerel\per\cubic\metre}~\cite{snolabhandbook}.
Approximately \SI{0.5}{\cubic\metre} of lab air was pumped through a NaOH trap to
remove CO$_2$, a cold trap at approximately \SI{-50}{\degreeCelsius} to remove water
vapour, and then a trap filled with a porous polymer medium (Chromosorb\textsuperscript{\textregistered}
102 manufactured by Supelco, a member of the Sigma Aldrich group) kept
at approximately \SI{-110}{\degreeCelsius} to capture radon. Oxygen, nitrogen and argon passed
through all traps. The trapped radon was concentrated in a small trap 
and then allowed to expand into a \SI{100}{\cubic\centi\metre} storage tube. 
This tube was then attached to the DEAP-1 argon system inlet, which was first pumped on to remove all contaminants and then filled with argon at a higher pressure than the circulation loop. 
The radon storage tube was opened to the argon inlet and then 
the argon was opened to the recirculation loop (through a SAES getter) 
until the pressure equilibrated. The inlet was then closed and 
re-pressurized. There were four injection cycles within a ten-minute period.

\begin{figure}[htbp] 
   \centering
   \includegraphics[width = \textwidth]{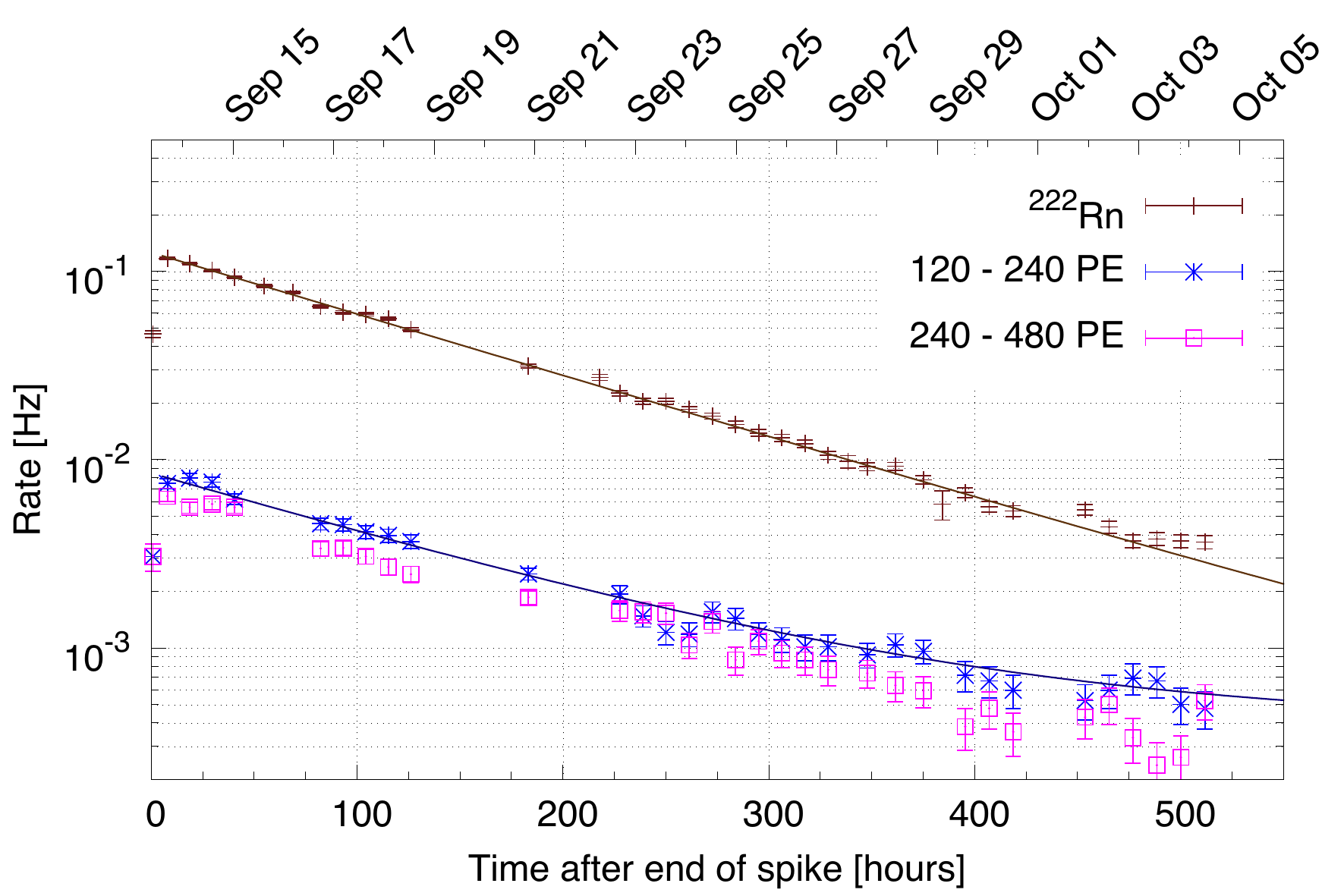}
   \caption{Radon alpha rate and event rate in two low energy windows (corresponding roughly to \SIrange{25}{50}{keVee} and \SIrange{50}{100}{keVee}) versus time after the radon spike. Expected decay curves based on the $^{222}$Rn lifetime and the constant rate before the spike are also drawn.}
   \label{fig:ratesovertime}
\end{figure}

The rates of the full-energy radon alphas and the population in two low-energy regions were tracked versus time as shown in
Fig.~\ref{fig:ratesovertime}. The event rate in the low-energy region decreases with a decay constant consistent with $^{222}$Rn, indicating that these low-energy events are in fact caused by radon decays.

The redesign of the detector chamber from G2 to G3, to include a neck plug and more tightly fitting acrylic
windows, was based on these findings. The change in rate of population B from G2 to G3 shows that the neck plug successfully removed many of the geometric backgrounds near the centre of the detector. 

The redesign of the acrylic windows had mixed results. Near the positive side (right) window, it reduced the geometric background rate. Near the negative side (left) window, the geometric background rate was very variable in time, at times increasing significantly. This indicates that the window was not placed tightly against the sleeve, and that the gap between the window and the sleeve was opening and closing. 

\subsubsection{Surface alpha backgrounds} \label{sect:surface}
The spectrum of events in population C in Fig.~\ref{fig:zfitvstotalpe} is consistent with the spectrum obtained from Monte Carlo simulation of radon-chain alpha decays on the TPB surface. 

\begin{figure}[htbp]
   \centering
	\includegraphics[width=\textwidth]{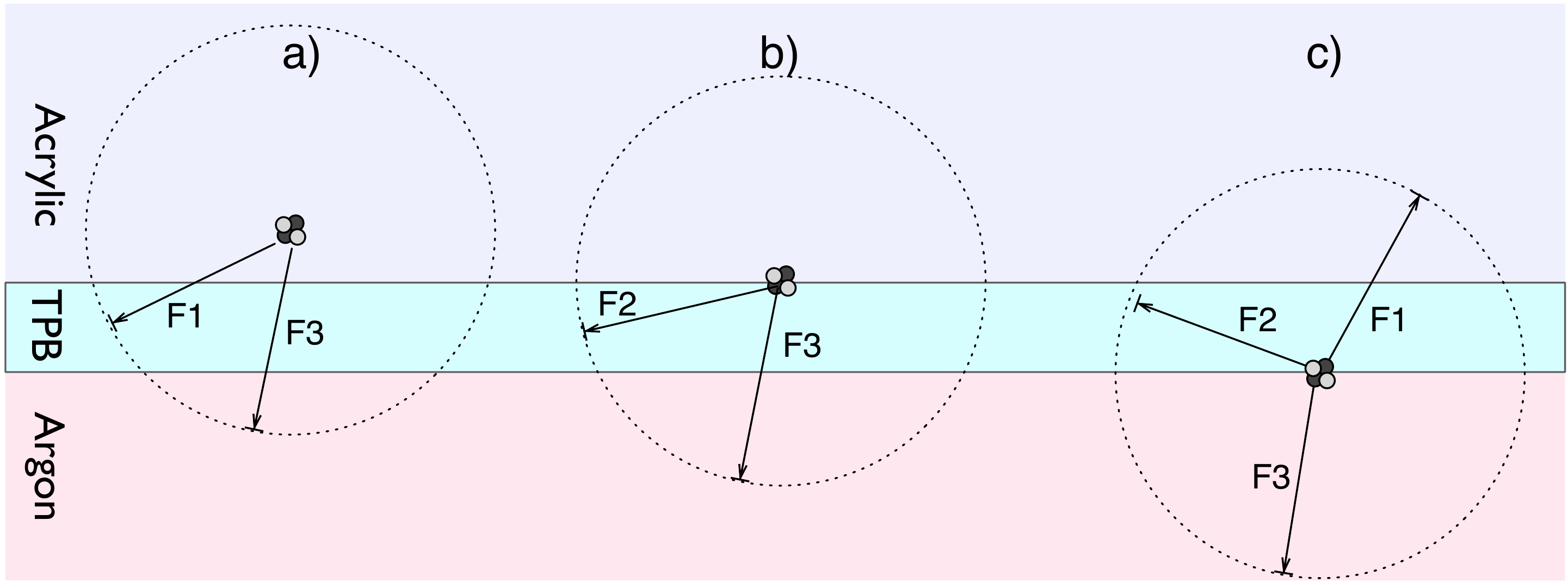}
   \caption{Sketch of possible paths that lead to light emission for an alpha particle emitted in the bulk acrylic (a), on the acrylic surface (b), and on the TPB inner surface (c).}
   \label{fig:alphapaths}
\end{figure}

The geometry of surface backgrounds in general is shown in
Fig.~\ref{fig:alphapaths}. Paths that lead to light emission are
labelled F1 to F3. The simulation returns the light generated by the alpha particle in both the liquid argon and the
TPB. The light yield of TPB under alpha particle
excitation was taken from Ref.~\cite{Pollmann:2011vf}. The light yield, which was determined at room temperature in that reference, was recently confirmed in an independent measurement and holds true at liquid argon temperature (87~K)\cite{Veloce:2013}.

The scintillation in
acrylic was assumed to be negligible, consistent with earlier measurements.

\begin{figure}[htbp]
   \centering
	\includegraphics[width=0.9\textwidth]{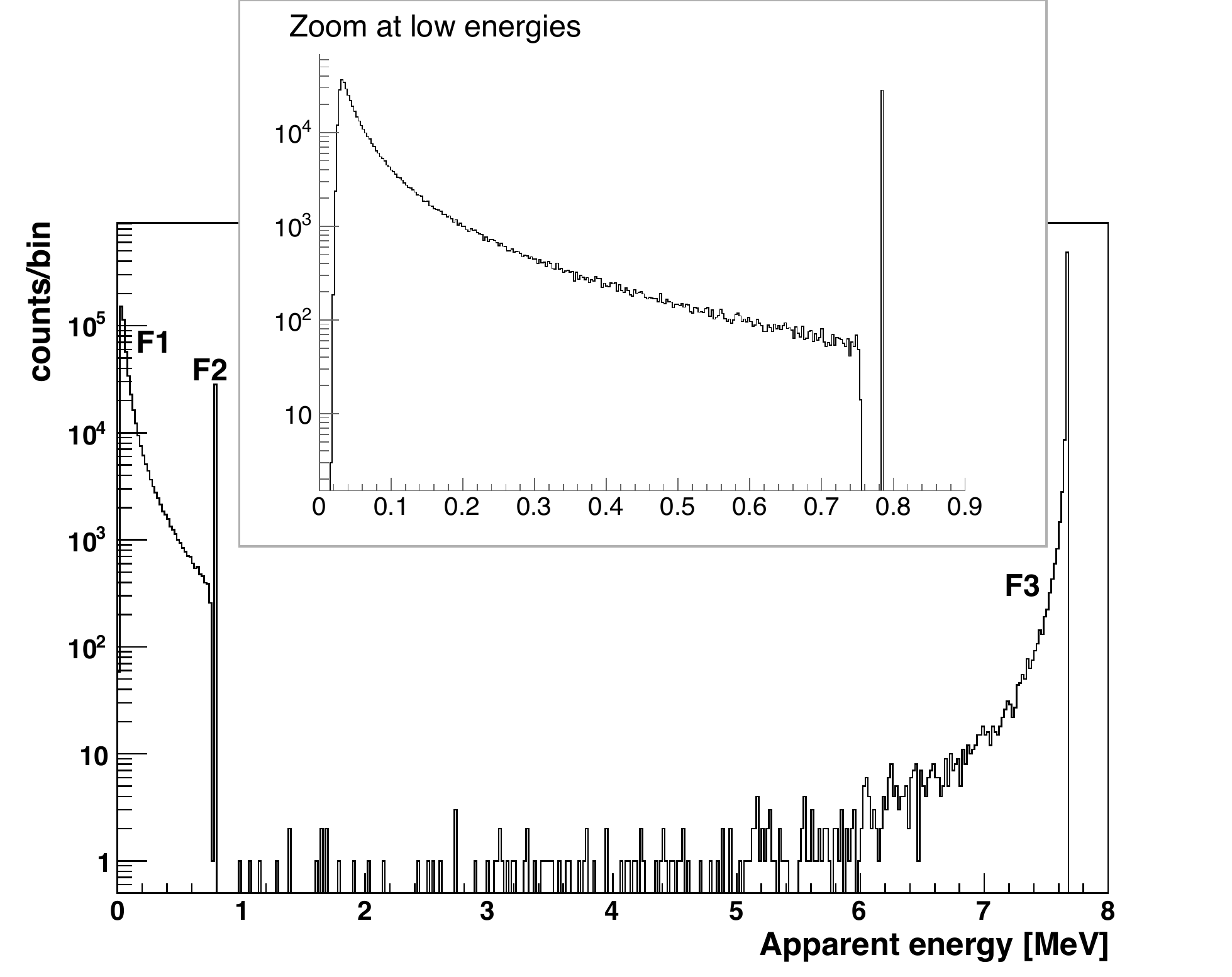}
   \caption{Simulated spectrum of \Nuc{Po}{214} alpha decays from the inner surface of the TPB (situation (c) in Fig.~\ref{fig:alphapaths}).}
   \label{fig:OnTPB}
\end{figure}

\begin{figure}[htbp]
   \centering
	\includegraphics[width=0.9\textwidth]{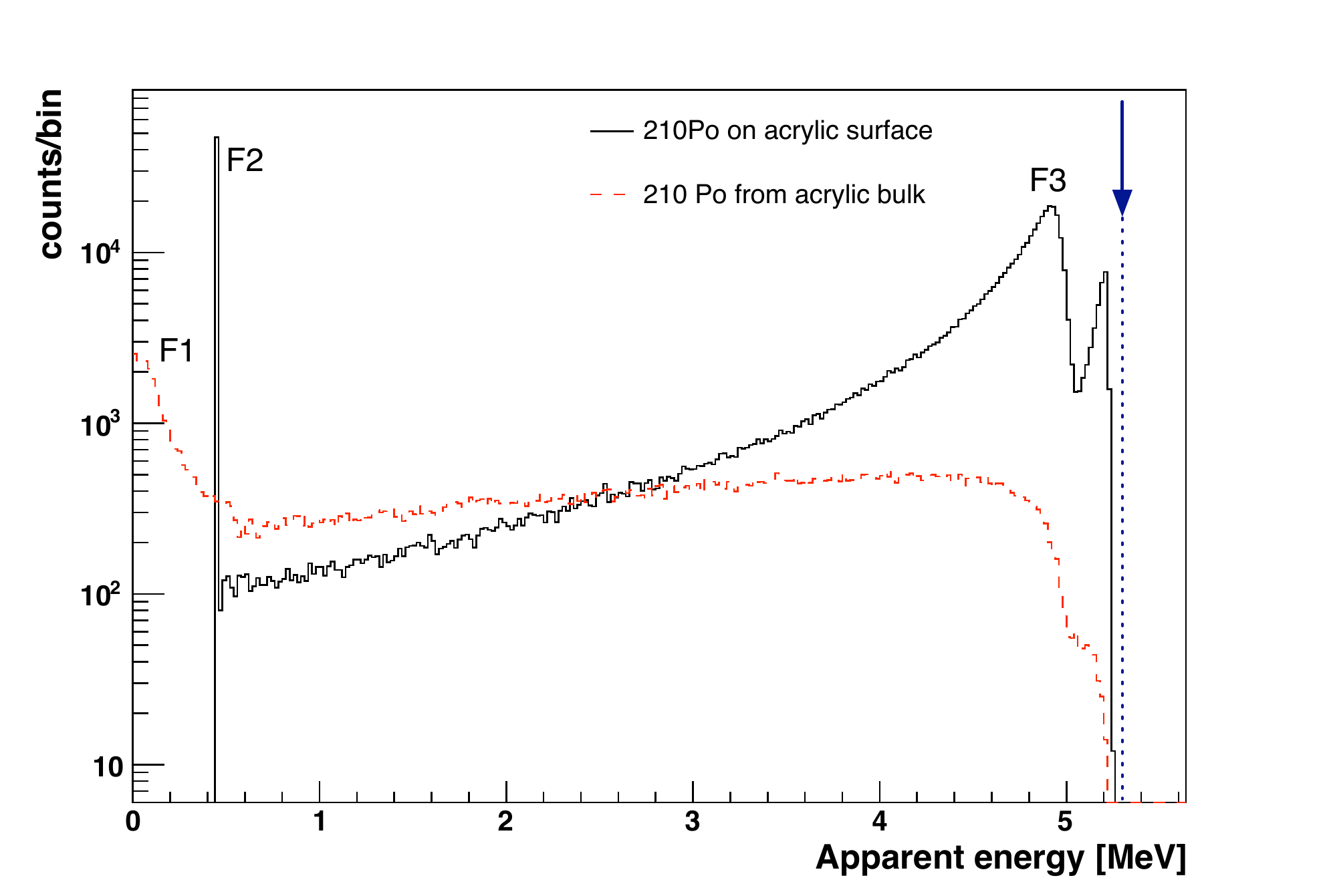}
   \caption{Simulated spectra of $^{210}$Po on the acrylic surface (situation (b) in Fig.~\ref{fig:alphapaths}) and in the bulk acrylic (situation (a) in Fig.~\ref{fig:alphapaths}). The full energy of the alpha particle is indicated by the arrow. }
   \label{fig:Pos}
\end{figure}

Spectra simulated with a SRIM~\cite{Ziegler20101818} based code, without detector resolution
effects, are shown in Fig.~\ref{fig:OnTPB} and Fig.~\ref{fig:Pos}. The
simulation includes the different TPB thicknesses on the acrylic
windows and sleeve (this is the reason for the double peak structure at F3 in Fig.~\ref{fig:Pos}). Features in the spectra are labeled with the same names as the
paths in Fig.~\ref{fig:alphapaths} that generate them.

We simulated spectra for all radon chain decays from the bulk acrylic, acrylic surface, bulk TPB and TPB surface, but only select ones are shown here.

The \Nuc{Po}{214} spectrum in Fig.~\ref{fig:OnTPB}, corresponding to
situation c) in Fig.~\ref{fig:alphapaths}, has a peak at low energies
(F1) at about 70~keVee. The lowest energy an event can have in this
case corresponds to the shortest path through the TPB. Any isotope
emitting a lower energy alpha will have a similar peak at a slightly
higher energy, since the energy loss increases for lower kinetic
energy particles. This means that given a smooth, uniform thickness TPB
layer and infinite energy resolution, there is a low energy cutoff
below which no such surface events occur.

A minimum visible energy also exists for alpha decays that happen on a perfectly smooth
acrylic surface.  The paths labeled F3 all end in the liquid argon,
so that most or all of the alpha's energy is visible. The lowest energy
event happens when the alpha loses all its energy in the TPB (F2
paths). For rough surfaces or when straggling is considered, events
with lower visible energy are possible.

The most problematic backgrounds are then from alpha decays in the acrylic
bulk or the TPB bulk, which lead to events down to 0~keV even under
the first order approximation of a smooth surface without straggling.

We have developed a detailed GEANT4 based model to study the combined effects
of surface roughness, contamination and straggling. It has been successfully
employed to re-interpret recent results by CRESST-II and suggested
a simple explanation for the reported excess of low energy events~\cite{KuAniak:2012ek}.
In the case of DEAP-1, for the TPB surface contamination,
no significant enhancement at the low energy end of the spectrum occurs, 
mainly because of the finite minimum thickness of the TPB coating, responsible for the low energy cutoff in the spectrum.
We have confirmed that for a realistic choice of TPB surface roughness parameters
informed by surface profile scans (a sinusoidal model with 20\% thickness variation and \SI{10}{\micro\metre} period waviness), 
the resulting spectra are well approximated by the ideal case as long as the coating is free of pinholes. 
This can be explained by the fact that the roughness scale is significantly smaller than the range of alpha particles.  

Sub-micron-resolution scans of TPB coatings and sanded acrylic, and investigation on possible effects of very small scale structures on both the low and high end of the energy spectrum, are planned.

Surface roughness plays a more significant role for $^{210}$Po activity on the acrylic surface and could potentially introduce
low energy tails in the spectrum. However, this possibility is excluded by the contamination limits discussed in the next section.

\subsubsection{Window backgrounds}
Event population (D) in Fig.~\ref{fig:zfitvstotalpe}, near both acrylic windows
at energies up to about \SI{100}{keV_{ee}}, was observed in every iteration of
the DEAP-1 detector but its origin is unclear. The geometric
background events from alpha particles emitted near the windows
reconstruct at a location slightly closer to the detector centre than
the window backgrounds, indicating that the window events might come
from the glass rather than the acrylic windows. 

In order to reduce the rate of these events, a 10~cm long region of the argon volume
centred on the middle of the detector is defined as the fiducial volume.

\section{Background rates and surface contamination}
\subsection{Limits on surface contamination by \Nuc{Pb}{210}} \label{sect:surfacebglimits}

Coating the inside of the G2 detector chamber with purified acrylic was meant to reduce backgrounds then presumed to be due to surface contamination. The failure of this coating to significantly reduce the background in the low energy region of interest, as well as the failure of the simulated surface background spectra to adequately describe the G1 and G2 data, motivated further studies, including the radon spike, that led to the identification of the geometric backgrounds. Geometric backgrounds dominated the background spectrum in G1 and G2, and they obscured a possible background reduction factor due to the purified coating.

The geometric backgrounds, at least near the centre of the detector, were sufficiently reduced in G3 to attribute the remaining events above about \SI{50}{keV_{ee}} in the nuclear recoil band to surface backgrounds. 
The high energy spectrum from G3 data (Fig.~\ref{fig:V5HighE}) was fit with the simulated
spectrum of $^{210}$Po on the acrylic (Fig.~\ref{fig:Pos} and
\ref{fig:V5HighE}) to determine the number of $^{210}$Po decays necessary to cause events around 5~MeV with the observed rate.  The
resulting contamination was found to be \SI{3.1d-8}{\becquerel\per\centi\metre\squared},
assuming all the events around \SI{5}{MeV} are due to this source. 

Events at intermediate energies between 1 and \SI{4}{MeV} near the centre of
the detector can come from $^{210}$Po on and in the acrylic. The energy
calibration at these energies has large uncertainties and the number
of events observed there is small (around 25). The spectrum was
therefore not fit, but the number of events integrated. Then the
fraction of $^{210}$Po decays in the acrylic that lead to events in
this energy range was determined from the simulation and used as a
correction factor on the observed number of events. Assuming that all
these events are from $^{210}$Po in the bulk acrylic, its concentration is
$(3.2\pm0.6)\cdot 10^{-5}$~Bq/cm$^{3}$ or $(1.6\pm 0.3)\cdot 10^{-19}
$~g/g. Based on this estimate, the number of $^{210}$Po events expected in the low energy 
part of the spectrum, i.e. between \SIrange{50}{200}{keV_{ee}}, is about two orders of magnitude below
the measured number of events.

These results can be interpreted as upper limits. Some geometric background events could
be contributing to the event rate around \SI{5}{MeV}, and some of the
alpha decays on the acrylic surface or in the TPB have energies between 1 and
\SI{4}{MeV}.

\subsection{Low energy background rates} \label{sect:lowenergyrates}
To illustrate how the background event rates declined in two typical low-energy regions, we show Table~\ref{tab:bgratet}. An Fprompt cut with (50 $\pm$1)\% nuclear recoil efficiency was applied.
\begin{table}[htbp]
   \centering
   \begin{tabular}{@{} lrr @{}} 
      \toprule
      	& \multicolumn{2}{c}{Background rates [\si{\micro\hertz}]} \\
Data set & 120-240~PE	& 50-100~keV$_{ee}$  \\
 \midrule
G1 NV &	 47 $\pm$4 & 47 $\pm$4 \\
G2 NV &	23 $\pm$3 & 30 $\pm$3 \\
G3 NV &	 10 $\pm$4  &  7 $\pm$4 \\
      \bottomrule
   \end{tabular}
   \caption{Background rates at (50 $\pm$1)\% nuclear recoil efficiency, along 10~cm of the detector corresponding to a fiducial volume of 2.5~kg. The rates are shown in the window of
TotalPE which is of interest for PSD analysis, as well as in a fixed window in keV$_{ee}$. The same TotalPE window translates into different energy windows because the three detector generations had
different light yields.}
   \label{tab:bgratet}
\end{table}

The full low-energy spectrum, and the origin of the remaining events, will be discussed in the next section.

\section{The low-energy background spectrum} \label{sect:lespectrum}
\begin{figure}[htbp]
   \centering
	\includegraphics[width=\textwidth]{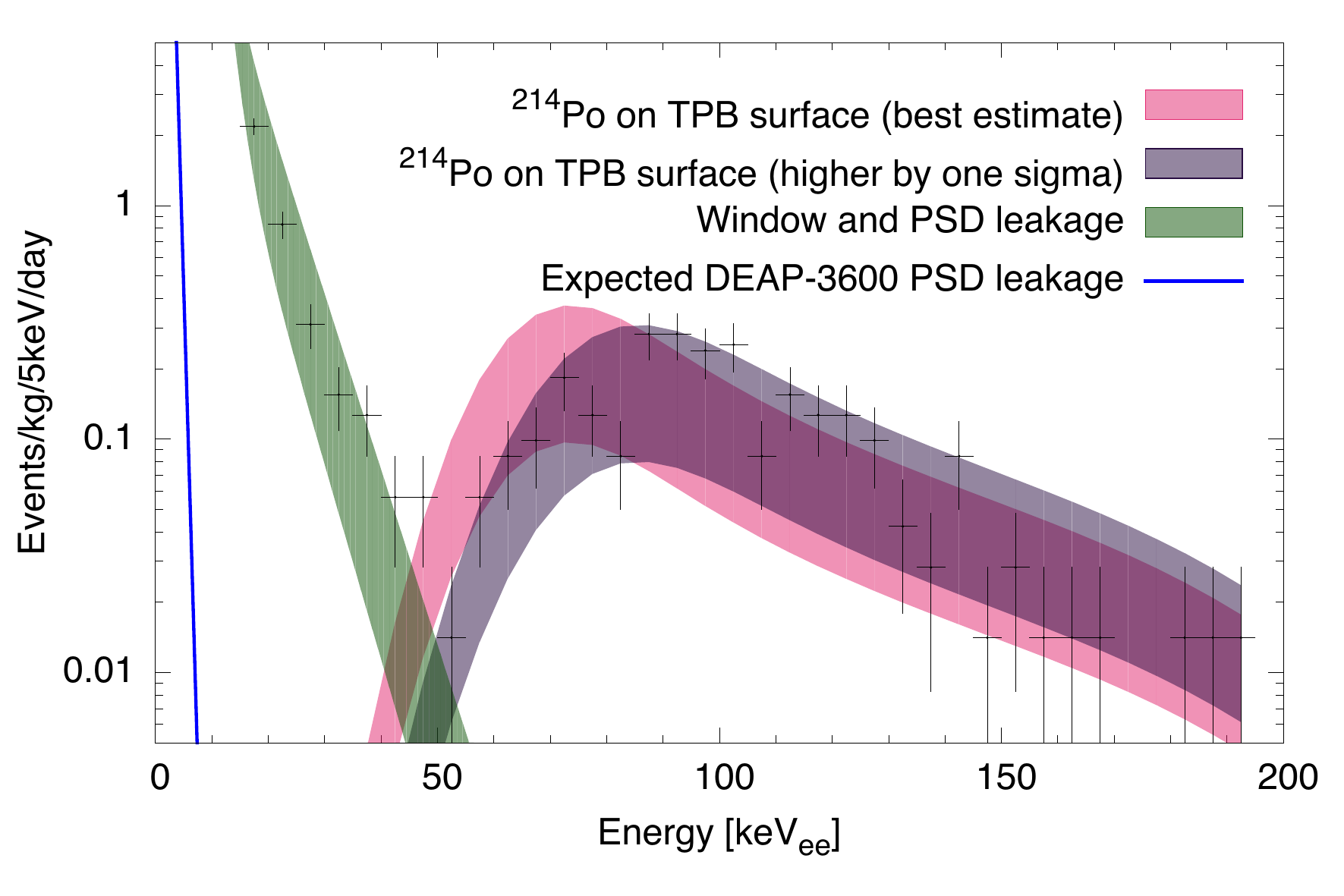}
   \caption{G3 low energy background spectrum with a PSD cut that has 85$\pm$5\% recoil efficiency. The estimated rate of window and PSD leakage into the data region is drawn in medium grey (green). The light grey (pink) band shows the expected location of the peak from \Nuc{Po}{214} on the TPB surface (compare Fig.~\ref{fig:OnTPB}), as a representative example for the alpha emitting radon chain isotopes, with best estimate parameters for the TPB thickness, TPB scintillation efficiency and nuclear recoil quenching factor in argon (alpha-emitting radon daughters other than \Nuc{Po}{214} also contribute to the measured peak, see text). The height in the y-direction of the shaded region corresponds to one sigma above and below the best estimate value for the rate.  The second, dark grey (purple) band indicates the one sigma upward uncertainty of the peak location in energy. The blue line shows the PSD leakage predicted for DEAP-3600.}
   \label{fig:V5spectrum}
\end{figure}

The low-energy spectrum of the lowest background detector configuration, G3 (NV), is shown in Fig.~\ref{fig:V5spectrum} for 85$\pm$5\% nuclear recoil efficiency. A Zfit cut of [-8 to 3], corresponding to the physical location of \SI{-5.0}{\centi\metre} to \SI{5.4}{\centi\metre}, was applied. 

Below \SI{50}{keV_{ee}}, the spectrum of Fig.~\ref{fig:V5spectrum} is well described by PSD leakage and window leakage.

PSD leakage refers to electromagnetic background events that the pulse
shape discrimination method misidentifies as nuclear-recoil events. The amount of PSD leakage is determined mainly by photoelectron statistics and thus increases exponentially towards lower
energies. The PSD leakage rate per energy bin was determined using dedicated PSD data runs with a
tagged $^{22}$Na source, which provided data up to about \SI{25}{keV_{ee}}, and it was extrapolated to higher energies from there.

Window leakage refers to events (from populations B and D) that should reconstruct on the windows
of the detector, but which, also due to low photo-electron statistics
at low energies, reconstruct in the fiducial volume. The amount of
leakage into the region of interest was extrapolated for each energy
bin by fitting the event position distribution with two gaussians near
the windows on top of a constant rate. The two gaussians were then
integrated over the region of interest to estimate how many window
events leak into this region. 

Both leakage contributions are included in the medium grey (green) band in Fig.~\ref{fig:V5spectrum}.

The events above \SI{50}{keV_{ee}} in the same figure are most likely from alpha decays on the TPB surface.
The light and dark grey (pink and purple) bands in Fig.~\ref{fig:V5spectrum} are the result
of the simulation described in Sect.~\ref{sect:surface}, for \Nuc{Po}{214} on
the TPB surface, with the argon scintillation from the recoil nucleus added.

\begin{figure}[htbp]
   \centering
	\includegraphics[width=\textwidth]{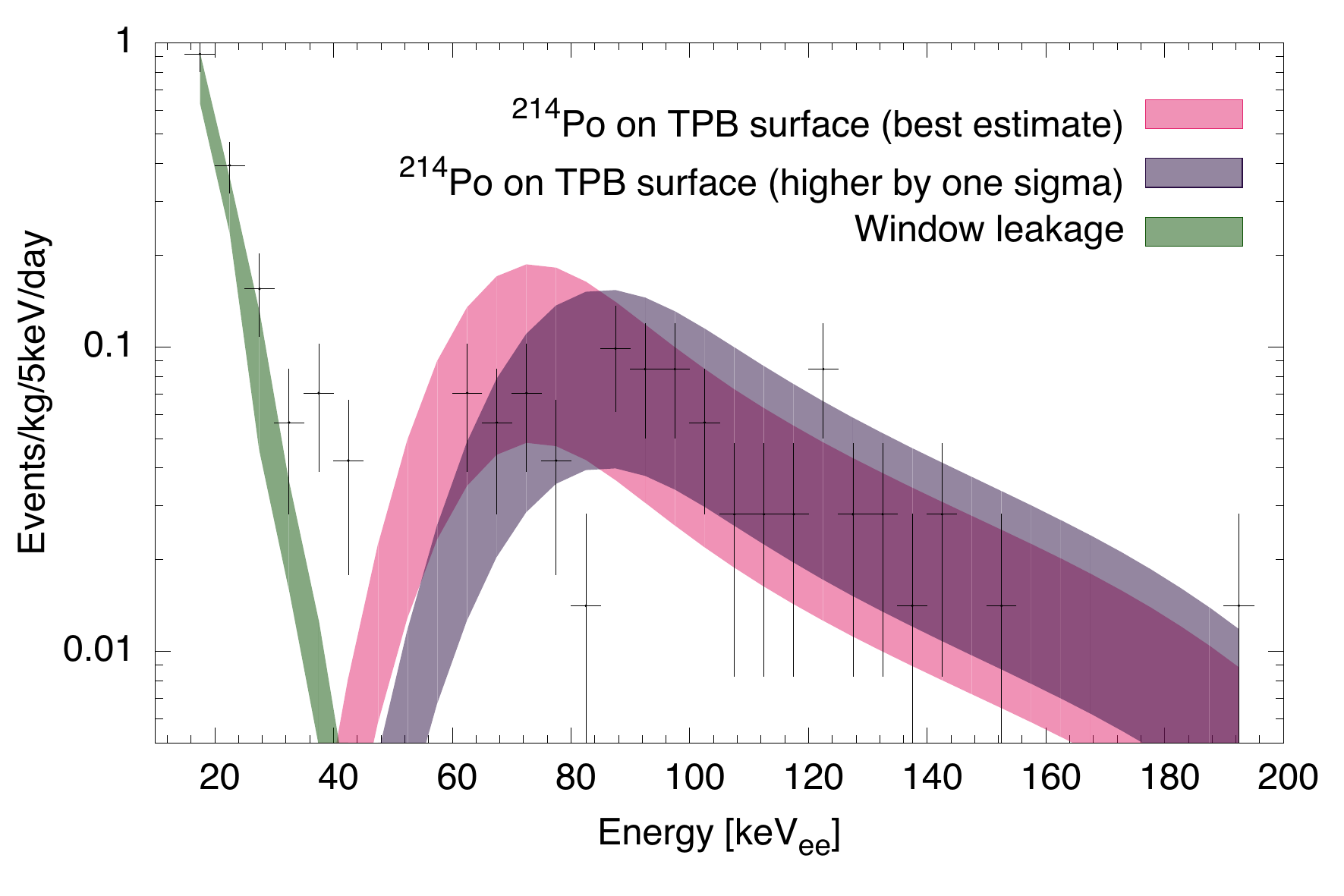}
   \caption{G3 low energy background spectrum. Same as Fig.~\ref{fig:V5spectrum}, but with 50$\pm$5\% recoil efficiency. Again, the expected spectrum of \Nuc{Po}{214} on the TPB surface is used as a representative example for the alpha emitting radon chain isotopes (alpha emitting radon daughters other than \Nuc{Po}{214} also contribute to the measured peak, see text).}
   \label{fig:V5spectrum50perc}
\end{figure}

As explained later in this section, the rate and location of the low energy peak is consistent with the low energy signature of radon chain alpha decays. The signatures of those are all nearly identical, with lower energy alpha decays producing peaks at slightly higher energies (because their energy loss through the thin TPB film is higher). Only a limited fraction of the low energy peak can be attributed to surface alpha decays of \Nuc{Po}{214} through Bi-Po tagging. However, as \Nuc{Po}{214} decays with the highest energy alpha, it makes the lowest energy TPB scintillation peak of the \Nuc{Rn}{222} chain daughters, thus representing the worst case in terms of the probability of an event leaking into the Dark Matter signal region. It is therefore used in this section, including Figures~\ref{fig:V5spectrum} and ~\ref{fig:V5spectrum50perc}, as a representative placeholder.

The normalization of the simulated spectrum is determined
by the number of \Nuc{Po}{214} events missing from the data at high energies (see Fig.~\ref{fig:V5HighE}) and the
efficiency of the cuts on Fprompt and Zfit. The uncertainty in the normalization is
dominated by the statistical uncertainty on the high energy alpha
rate. Even though the events at \SIrange{50}{200}{keV_{ee}} in the low energy spectrum are not all from the plated out \Nuc{Po}{214}, the normalization of the simulated spectrum is at the right order of magnitude for events from the radon chain. 


The energy of each event in the peak of the simulated spectrum includes a contribution from the TPB induced alpha scintillation and from the scintillation of argon due to the recoil nucleus. The peak location depends on the TPB thickness, TPB scintillation yield ($\epsilon_{tpb}$) for alphas
and the quenching factor for nuclear recoils in argon ($q$). 
The simulation returns the energy deposited in TPB ($E_{tpb}$), from which the amount of scintillation light that the TPB emits can be calculated as $E_{tpb}\cdot \epsilon_{tpb}$. An event with that amount of light will be seen as having an energy of $\frac{E_{tpb}\epsilon_{tpb} }{ \epsilon_{lar}}$, where $\epsilon_{lar}$ is the liquid argon scintillation yield for electron-recoil events, because that is how we calibrate the energy scale in DEAP-1.
The total apparent event energy for a polonium decay on the TPB surface in keV$_{ee}$ is then
\begin{equation}
E_{app} = \frac{E_{tpb}\epsilon_{tpb} }{ \epsilon_{lar}} + q\cdot E_{r,lar}
\end{equation}
where $E_{r,lar}$ is the energy of the recoil nucleus going into the liquid argon.
The experiment detailed in Ref.~\cite{Pollmann:2011vf} was done with the same type of PMT and similar DAQ setup as was used in DEAP-1 G1. In this experiment we found that $\epsilon_{tpb} = 0.20 \pm 0.05$~PE/keV. The DEAP-1 G1 light yield was $\epsilon_{lar} = 2.4 \pm 0.2$~PE/keV. We assumed the fraction $\frac{\epsilon_{tpb}}{\epsilon_{lar}}$ to remain constant, since we think most of the light yield increase in later generations of DEAP-1 was due to the better PMTs and electronics, which would affect both values the same. For the quenching factor we used $q = 0.25 \pm 0.03$~\cite{Gastler:2012ki}.
The light gray (pink) band shows $E_{app}$ for each simulated event using the central values for each parameter. The one sigma uncertainty in one direction on these parameters ($E_{app} + \Delta E_{app}$ ) is indicated by the dark grey (purple) band in Fig.~\ref{fig:V5spectrum}.

The spectrum for 50$\pm$5\% recoil efficiency is shown in Fig.~\ref{fig:V5spectrum50perc}. The medium grey (green) band includes only window event leakage, since not enough data was available to extrapolate the gamma leakage here.

For high energy alpha events from \Nuc{Po}{214}, we expected 64\%\ Bi-Po tagging efficiency and observed a beta particle from \Nuc{Bi}{214} preceding 79$\pm$19\% of them (see Sect.~\ref{sec:tagging}). Likewise, we expect that a similar fraction of the \Nuc{Po}{214} events in the spectrum between \SIrange{50}{200}{keV_{ee}} should be preceded by an electron.

We find that no more than 7\% of the events in the peak are preceded by an electron within a timespan of three times the \Nuc{Po}{214}'s half-life.  Taking into account the expected Bi-Po tagging efficiency of 64\%, we conclude that only up to 11\% of the events in the peak in Fig.~\ref{fig:V5spectrum} are from \Nuc{Po}{214}. As discussed in Sect.~\ref{sec:bipos}, the rest of the \Nuc{Po}{214} events expected at low energies are closer to the windows, and thus not included in this data.

The remaining events could in general be from any of the radon daughters decaying on the TPB surface.

If some of the \Nuc{Rn}{222} and \Nuc{Po}{218} are frozen out on the TPB surface, alpha particles from their decays going into the TPB can contribute events to the low energy peak.

A full energy event is seen if a decay happens in the bulk argon, or if the decay happens on the TPB surface and the alpha particle is emitted into the argon. The number of full energy events, $N_{\text{HE}}$, in general can thus be written as
\begin{equation}
N_{\text{HE}} = N_{\text{Rn}} \sum_x [P^{x}_A + (1 - P^{x}_A)\cdot 1/\hat{\eta} ]
\end{equation}
where $N_{\text{Rn}}$ is the number of \Nuc{Rn}{222} decays, $P_{A}$ is the probability for an isotope to be in the bulk liquid argon, and $1/\hat{\eta}$ is the probability for an isotope on the TPB surface to emit the alpha particle into the liquid argon. $\hat{\eta}$ is equal to the $\eta$ from Table~\ref{tab:spectrafit} if all \Nuc{Po}{214} is on the TPB surface, and under the assumption that loss of isotopes due to their drifting out of the active volume is negligible.
The sum runs over $^{222}$Rn, $^{218}$Po and \Nuc{Po}{214}, and in general also over the isotopes in the $^{220}$Rn chain, though their overall rate is lower by a factor of about ten and therefore negligible. As discussed in Sect.~\ref{subs:rnspectra}, the rate of $^{210}$Po decays is strongly suppressed, and contributes only a couple of events to the spectrum. 

The number of events that could then contribute to the peak in the low energy spectrum is
\begin{equation}
N_{\text{LE}} = N_{\text{Rn}} \sum_x [(1-P^{x}_A)(1-1/\hat{\eta} )]
\end{equation}
where $N_{\text{LE}}$ is the number of events from \SIrange{50}{200}{keV_{ee}} in Fig.~\ref{fig:V5spectrum}, corrected for cut efficiencies.

With our detector, we cannot measure enough of the parameters to determine a unique solution to those equations. If we make the simplifying assumption that $P_{A}^{Po214}$ is zero and that $P_{A}^{Rn222} = P_A^{Po218}$, solutions consistent with our observations of low energy and high energy event rates and radon tag rates have $P_{A}^{Rn222}$ in the range of 65\% to 75\%.

\section{\Nuc{Po}{214} decays at high and low energies}\label{sec:bipos}

The full energy alpha spectra show that the \Nuc{Po}{214} peak is reduced compared to the \Nuc{Rn}{222} peak, necessitating the introduction of a scaling factor in the fit. \Nuc{Po}{214} decays can evade detection at their expected energy if the parent isotope drifted outside the active volume, or by emitting the alpha particle into the TPB layer, in which case the events appear at much lower energies in the spectrum. Because \Nuc{Po}{214} can be identified using the Bi-Po coincidence tag (see Sect.~\ref{sec:tagging}) regardless of their apparent energy, we should be able to find the majority of \Nuc{Po}{214} decays that take place in the active liquid argon volume or on a TPB surface. 

Because of the large surface area of the windows, and the likelihood that \Nuc{Po}{214} is plated out on surfaces, the whole detector volume has to be taken into account when looking for all \Nuc{Po}{214} decays. At high energies, simple cuts on energy and Fprompt, without restrictions on Zfit, are sufficient to select a population of alpha events on which to base the Bi-Po coincidence analysis without significant contributions from other background sources. 

At low energies, the high rate of events at the windows, which is not related to radon in the active volume, poses a twofold problem: they introduce both a larger fraction of random coincidences compared to real Bi-Po coincidences, and real Bi-Po coincidences not relevant to this analysis. The former issue is because the event population selected for tagging contains a smaller fraction of \Nuc{Po}{214} alphas. The latter issue, which is particularly problematic, arises because some of the low-energy events near the windows of the detector are ``geometric backgrounds'', i.e. degraded radon chain alphas (see Sect.~\ref{susbs:geometric}). Although a certain fraction of these are \Nuc{Po}{214} alphas, preceded by degraded \Nuc{Bi}{214} betas, they are irrelevant for comparison with the high-energy Bi-Po coincidences and have to be excluded. 

The analysis of the low energy background presented in Sect.~\ref{sect:lespectrum} includes only those low-energy nuclear-recoil-like events that occur near the centre of the detector, which renders the contribution of backgrounds associated with the windows negligible. Here, by discarding runs with the highest rate of window events and keeping the remaining 73\%\ of the data-set used in the rest of this work, we can relax the fiducial volume cut without increasing the contribution of window events too much (in DEAP-1 G3, the rate of window events fluctuated significantly, see end of Sect.~\ref{susbs:geometric}). We then require that the spectrum of electron-recoil events preceding the population of low-energy nuclear-recoil-like events selected for tagging be consistent with the nearly background-free spectrum of electron-recoil events preceding the high-energy alpha events in the 6200--8600~PE range, which corresponds to the peak energy of \Nuc{Po}{214} alphas. (No Bi-Po coincidences above random expectation were observed between 5000 and 6200~PE.) The Kolmogorov-Smirnov test was used to check if the unbinned energies in both populations of electron-recoil events came from the same parent energy distribution. 

The Zfit and energy cut selecting the population of low-energy nuclear-recoil-like events to be tagged was progressively relaxed, and in each step the spectrum of preceding electron-recoil events evaluated, as long as the electron-recoil event spectra proceeding the low and high energy nuclear-recoil events were consistent to a significance level of $\alpha$=0.05.

\begin{figure}[htpb]
   \centering
	\includegraphics[width=\textwidth]{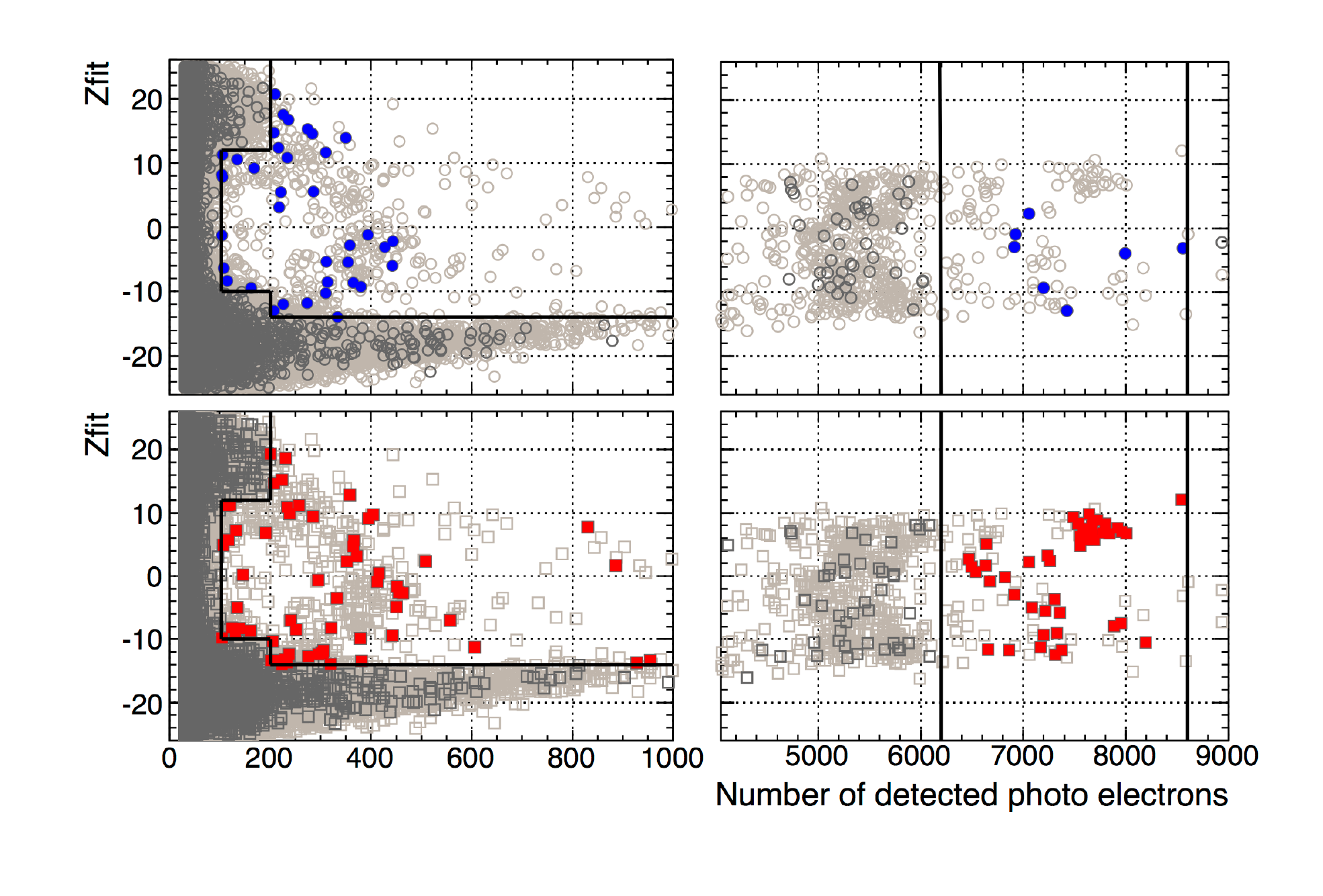}
   \caption{Bi-Po coincidences in a subset of G3 NV data. The plots show Zfit vs the total number of photo electrons for high Fprompt events, drawn in light grey, in low-energy (left) and high-energy (right) regimes. On the upper two panels, events that are followed by a beta event within 3 half-lives of \Nuc{Po}{214} are drawn in dark grey. The subset of those that is accepted by the optimized selection cuts (indicated by the black bars, also see text) is highlighted in blue. 
   On the lower two panels, events that are preceded by a beta within 3 half-lives of \Nuc{Po}{214} are drawn in dark grey, and the subset of those accepted by the optimized cuts is highlighted in red.}
   \label{fig:bipopanels}
\end{figure}

We thus found that the contribution of geometric \Nuc{Po}{214} alpha events and possibly other backgrounds was negligible for events in the 200--1000~PE (\SIrange{50}{250}{keV_{ee}}) range and Zfit between -14 and 25, and for events in the 100--200~PE (\SIrange{25}{50}{keV_{ee}}) range and Zfit between -10 and 12 (see Fig.~\ref{fig:bipopanels}). At lower TotalPE and further out from the centre of the detector, the beta spectra are no longer consistent.

\begin{figure}[htpb]
   \centering
   \begin{overpic}[width=\textwidth]{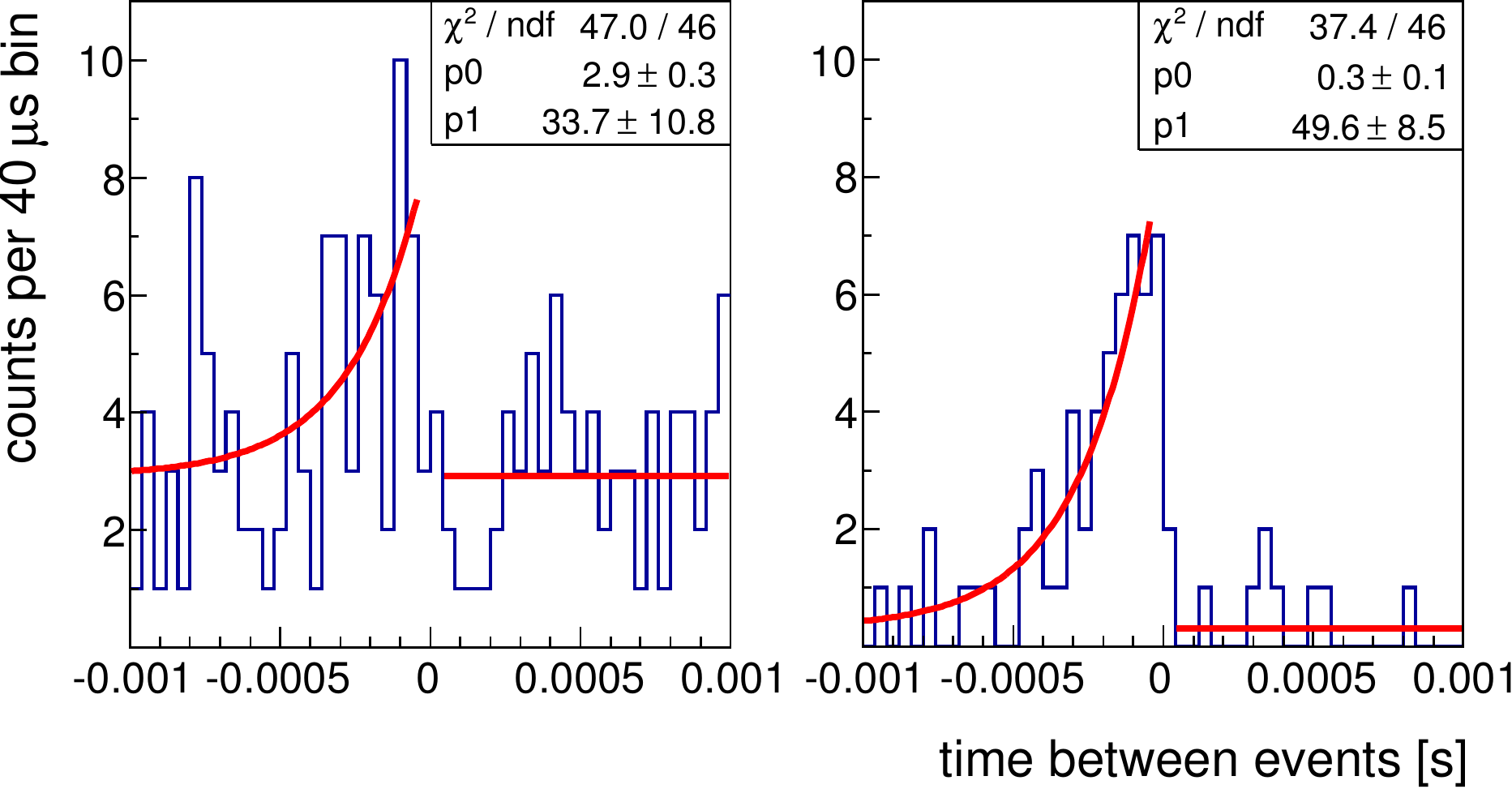}
   \put(11,47){\small low energy}\put(59,47){\small high energy}
   \end{overpic}
   \caption{Histogram of the time difference, $\Delta t$, between the low-energy (left panel) and high-energy (right panel) alpha events selected by the optimized cuts (see text and Fig.~\ref{fig:bipopanels}) and the beta events preceding and following them. In both energy regimes, an excess of preceding betas is seen, consistent with the decay of \Nuc{Po}{214}.  The fit function is given by $p_0+H(-t)\frac{p_1}{\tau}\exp\left(\frac{\Delta t}{\tau}\right)$, where $H$ denotes the Heaviside step function and $\tau$ is fixed to the mean \Nuc{Po}{214} life-time of \SI{237}{\micro\second}. Two bins next to the time difference ``0'' are prone to dead time effects and are excluded from the fit.}
   \label{fig:bipotime}
\end{figure}

In order to find the rate of real Bi-Po coincidences in the data-set restricted to the largest Zfit range possible while excluding contributions from other background sources, the histogram of the relative times between the nuclear-recoil events, and both the preceding and following electron-recoil events, was fit with an exponential function plus a constant. The \Nuc{Po}{214} half-life was fixed to the literature value, and the timing spectrum of following events was used to determine the rate of random coincidences, as shown in Fig.~\ref{fig:bipotime}.

Some low-energy Bi-Po coincidences can still be missed, even with the relaxed Zfit and energy cuts. In order to account for that systematically, their rate at Zfit$>$10 was estimated to be 3.4$\pm$\SI{2.4}{\micro Bq} using the fitting technique described above. A similar rate is expected at the negative Zfit end of the chamber, i.e for Zfit$<$-14, assuming that the true events are similarly distributed. As a correction, \SI{3}{\micro Bq} are added to the final rate of Bi-Po's at low energies and the uncertainty is increased by the same value. We neglect the inefficiency due to the energy cut (which we expect to be small in comparison with the global uncertainty).

In Tab.~\ref{tab:bipos} the resulting rates for high- and low-energy nuclear-recoil-event populations are summarized and compared with rates of tagged high-energy \Nuc{Rn}{222} events and inferred \Nuc{Po}{214} events.
\begin{table}[htbp]
   \centering
   \begin{tabular}{@{} cccc @{}} 
      \toprule
      \Nuc{Rn}{222} [\SI{}{\micro Bq}] & HE Po [\SI{}{\micro Bq}] & HE Bi-Po [\SI{}{\micro Bq}] & LE Bi-Po [\SI{}{\micro Bq}] \\
 \midrule
 99$\pm$15 & 36$\pm$6 & 29$\pm$5 & 22$\pm$10 \\
      \bottomrule
   \end{tabular}
   \caption{Comparison of event rates in G3 data (NV, restricted to 73\% of the data set). Rates of, respectively: \Nuc{Rn}{222} decays (from alpha tagging, see Sect.~\ref{sec:tagging}, corrected for a tagging efficiency of 82$\pm$9\%), \Nuc{Po}{214} decays at high energies (rate of events in the full energy alpha peak, inferred from the total number of high energy alpha events in NV data, assuming that the fraction of \Nuc{Po}{214} events in the high energy spectrum is the same as in the LV data~\ref{subs:rnspectra}), Bi-Po coincidences with high-energy events, Bi-Po coincidences with low-energy events (corrected for the Zfit cut efficiency, see text).}
   \label{tab:bipos}
\end{table}

The comparison presented in Tab.~\ref{tab:bipos} combines in a non-trivial way information from three different analysis types: \Nuc{Rn}{222}-\Nuc{Po}{218} tagging, global fit to the high-energy spectrum from LV data, and Bi-Po tagging. The fitted LV data high-energy spectrum provides the empirical efficiency correction for the \Nuc{Rn}{222} tag and the scaling factor, $\eta$, which allows us to derive the rate of \Nuc{Po}{214} alphas from the rate of \Nuc{Rn}{222} alphas (direct fit to NV data is not feasible due to poor energy resolution and PMT non-linearity).

As a basic self-consistency check and an indication that the properties of the detector (scaling factor $\eta$, and the \Nuc{Rn}{222} tag efficiency) remained the same within uncertainties in both the LV and NV data-sets we note that the ratio of \Nuc{Rn}{222} (from tagging) and \Nuc{Po}{214} alpha rates (inferred from total HE rate) is 2.7$\pm$0.7, which is consistent with $\eta$=4.0$\pm$0.7 measured in the G3 LV data.

We see the following evidence that \Nuc{Bi}{214} and \Nuc{Po}{214} decays occur on surface:
\begin{itemize}
\item The rates of high-energy and low-energy Bi-Po's are consistent (for surface decays half of the \Nuc{Po}{214} decays manifest as high- and the other half as low-energy events, see Sect.~\ref{sec:tagging}-\ref{subs:rnspectra}).
\item The Bi-Po tagging efficiency for high-energy events equals 79$\pm$19\%, which is consistent with the expectation of 64\%\ for surface decays (see Sect.~\ref{sec:tagging}).
\end{itemize}

\section{Discussion} \label{sect:discussion}
Radon and its decay products cause three distinct types of alpha
backgrounds. If the decay happens in the active liquid argon volume, a
full energy event is observed which can be readily discriminated from
a WIMP event based on its energy. A sustained full energy radon rate
between 110~and~\SI{180}{\micro\becquerel} (or 16 to \SI{26}{\micro\becquerel\per\kilogram})  for $^{222}$Rn and between 15~and~\SI{25}{\micro\becquerel} for $^{220}$Rn was observed in each iteration of DEAP-1. The radon is
believed to come from the the process system and the glass windows.

Some of the alpha decays occur in liquid argon volumes outside
the active volume, near the windows and the neck of the detector, from
where only a fraction of the scintillation light can reach the PMTs. These
geometric alpha events have an energy spectrum reaching down to
the energy threshold of the detector. A $^{222}$Rn spike demonstrated
the relation between low energy backgrounds and radon rate, and
modification of the detector to reduce the gap sizes near the window
and neck resulted in reduced low energy backgrounds.

Bi-Po tagging analysis (Sect.~\ref{sec:bipos}) provided evidence that longer lived \Nuc{Rn}{222} daughters, which precede \Nuc{Po}{214} in the radioactive
decay chain, plate out on inner surfaces of the detector. 

Backgrounds due to decays of radon and its daughter products in or on the TPB and the acrylic were studied in detail by means of Monte Carlo simulations. 

When radon daughter products decay on the TPB layer, they can emit an alpha particle into the TPB while the
recoil nucleus goes into the liquid argon, resulting in events with visible energies
between \SIrange{60}{200}{keV_{ee}}. Given an even TPB coating, the combined visible energy from the recoil nucleus in the liquid argon and the alpha particle in the TPB is never lower than this range.
 
Out of equilibrium \Nuc{Po}{210} is expected to be present in and on the
acrylic because the beads acrylic is typically made from are saturated
in \Nuc{Rn}{222}, and the acrylic chamber was in contact with air for a
few minutes during production. The decay of \Nuc{Po}{210} on the acrylic surface
and in the acrylic bulk was thus simulated, and limits for contamination of the DEAP-1 bulk acrylic and acrylic surface with \Nuc{Po}{210} of $(1.6\pm 0.3)\cdot 10^{-19}
$~g/g and \SI{3.1d-8}{\becquerel\per\centi\metre\squared} were obtained. The expected contribution
of this background to the low energy spectrum,
based on the rate observed at intermediate and high energies, is two
orders of magnitude below the observed background rates.

The events in the low energy background spectrum of DEAP-1 G3 can be explained by PSD and window leakage up to about 50~keVee. Between 50 and 200~keVee, a population of events from surface alpha decays is expected, and the number of events observed there is consistent with the radon contamination of the detector.

The higher projected light yield and improved electronic noise level in DEAP-3600 will strongly reduce the
PSD leakage,
while the simpler geometry is expected to eliminate the window
events. The medium grey (green) band in Fig.~\ref{fig:V5spectrum} will thus be shifted to the blue line shown in the same figure.
The surface alpha events are at the upper end of the \SIrange{20}{40}{keV_{ee}} energy region of interest in DEAP-3600. They will cause less background in the region of interest than the design goal if the overall radon rate in that detector's argon is kept below \SI{0.1}{\milli\becquerel}. The radon level in DEAP-3600 is expected to be ten times lower than this due to the use of materials with a low radon-emanation rate and the use of a radon trap in the DEAP-3600 process systems.

\section{Conclusions}

Window leakage, PSD leakage, geometric alpha events, and surface alpha events, combined, form a model that is consistent with the low energy background event spectrum observed in DEAP-1

The event class leading to window leakage is not expected to exist in DEAP-3600 due to the different detector
geometry. PSD leakage in DEAP-3600 will be strongly suppressed with higher light yield and less electronic noise than present in DEAP-1. The neck in DEAP-3600, being the only place where geometric alpha events can occur, was designed to strongly suppress their rate, and the planned argon purity with regards to radon will keep the surface alpha event rate significantly below the required level. 

This work suggests that the planned purity, with respect to radon, of the liquid argon for DEAP-3600 is more than adequate to suppress background events from radon chain isotopes introduced with the argon.

Backgrounds caused by radioactive radon daughter isotopes embedded in acrylic were found not to be dominant in DEAP-1. The DEAP-1 acrylic is radio-pure enough that, were it used for DEAP-3600 detector, the background target could already be reached, even when assuming a very conservative position-reconstruction resolution, albeit only with a reduced target mass. The acrylic used in the DEAP-3600 detector is expected to be significantly more radio-pure due to a strict quality assurance and assay program, and acrylic surface contamination will additionally be removed by in-situ resurfacing of the inner acrylic surface (for details see~\cite{2011AIPC.1338..137C}). 

This suggests that with the current design and radio-purity specifications for DEAP-3600 we can meet the background target of less than 0.6 event in 3 tonne-years. The detailed background budget for DEAP-3600 is presented in Ref.~\cite{DEAPTaup2012}.

\section{Acknowledgements}
This work has been supported by the Canada Foundation for Innovation (CFI), the Natural Sciences and Engineering Research Council of Canada (NSERC) and the Ontario Ministry of Research and Innovation (MRI).
The High Performance Computing Virtual Laboratory (HPCVL) has provided us with CPU time, data storage and support. We would also like to thank David Bearse for invaluable technical support and the SNOLAB staff for on-site support. The work of our co-op and summer students is gratefully acknowledged.

\bibliographystyle{elsart-num}
\bibliography{library}

\end{document}